\documentclass{article}
\usepackage[english]{babel}
\usepackage{subfiles}
\usepackage[letterpaper,top=2cm,bottom=2cm,left=3cm,right=3cm,marginparwidth=1.75cm]{geometry}

\usepackage{amsmath}
\usepackage{mathdots}
\usepackage{graphicx}
\usepackage[colorlinks=true, allcolors=blue]{hyperref}
\usepackage{amsthm}
\usepackage{tikz}
\usepackage{amssymb}

\usepackage{quantikz}

\usepackage{authblk}
\newtheorem{Def}{Definition}

\newtheorem*{Pre*}{Proof}

\newtheorem{theorem}{Theorem}[section]

\newtheorem{lemma}[theorem]{Lemma}
\newtheorem*{remark}{Remark}

\title{Fast Laplace transforms on quantum computers}

\author{Julien Zylberman}
\affil{Sorbonne Université, Observatoire de Paris, Université PSL, CNRS, LERMA, F-75005 Paris, France}
\date{}

\begin{document}
\maketitle

\begin{abstract}

While many classical algorithms rely on Laplace transforms, it has remained an open question whether these operations could be implemented efficiently on quantum computers. In this work, we introduce the Quantum Laplace Transform (QLT), which enables the implementation of $N\times N$ discrete Laplace transforms on quantum states encoded in $\lceil \log_2(N)\rceil$-qubits. In many cases, the associated quantum circuits have a depth that scales with $N$ as $O(\log(\log(N)))$ and a size that scales as $O(\log(N))$, requiring exponentially fewer operations and double-exponentially less computational time than their classical counterparts. These efficient scalings open the possibility of developing a new class of quantum algorithms based on Laplace transforms, with potential applications in physics, engineering, chemistry, machine learning, and finance.

\end{abstract}

\section{Introduction}

The Laplace transform is attributed to the French mathematician Pierre-Simon, Marquis de Laplace, for his work on the theory of probability published in 1812 \cite{simon1812theorie}. Laplace introduced a discrete version of the transform, related to what is called today the Z-transform, to study generating functions. After more than two centuries, the Laplace transform and its variations are widely used across all domains of science, particularly in physics \cite{bateman1910solution,doetsch2012introduction,mclachlan2014laplace,goswami2018numerical}, engineering \cite{doetsch2012introduction,mclachlan2014laplace,sawant2018applications}, chemistry \cite{davies1986testing,dobaczewski2004laplace,montella2007numerical,wojcik2015application}, biology \cite{membrez1996use,caputo2019inverse}, environmental science \cite{enting2007laplace}, cryptography \cite{hiwarekar2013application,genccouglu2017cryptanalysis}, signal processing \cite{jury1964theory,palani2022z}, machine learning \cite{smola2008introduction,szymczyk2015supervised,goodfellow20166} and economics \cite{grubbstrom1967application,daci2018application,daci2019laplace}. One notable example is its use in the resolution of integro-differential equations where derivatives and integrals are transformed into multiplication and division. Additionally, convolutions are simplified into products, reducing integro-differential equations to simpler polynomial equations. 

With the emergence of digital computers in the twentieth century, numerical computations were largely developed and many algorithms were introduced to evaluate Laplace transforms and their inverses \cite{titus1955general,soli1969laplace,dubner1968numerical,lego1959general}. Brute force algorithms transform a vector of $N$ components into one of $M$ components through a discrete Laplace transform with complexity $O(NM)$. This scaling was improved to $O(N+M)$ using various approximation techniques, maintaining logarithmic complexity with respect to the inverse of the error $1/\epsilon$ \cite{rokhlin1988fast,strain1992fast,loh2023fast}.

Nowadays, new computers are being developed leveraging the quantum properties of matter to perform computations differently. Due to superposition and entanglement, quantum computers have the potential to solve some problems with exponential or polynomial speedup compared to classical computers. For instance, factoring, computing discrete logarithms, and simulating dynamical systems can exhibit exponential speedups \cite{shor1994algorithms,shor1999polynomial,babbush2023exponential}. One common point of these algorithms is the Quantum Fourier Transform (QFT) which enables the implementation of the discrete Fourier transform of a vector of size $N$ with only $O(\log(N)^2)$ primitive quantum gates \cite{coppersmith2002approximate}. This requires a superpolynomially smaller number of operations compared to the classical Fast Fourier Transform which has complexity $O(N\log(N))$ \cite{cooley1965algorithm}.  The implementation of other fundamental transforms has remained an open problem for more than ten years \cite{Craigblog}.

In this work, we introduce the Quantum Laplace Transform (QLT), which enables the implementation of $N\times N$ discrete Laplace transforms on gate-based quantum computers.  The main idea is to expand each term of the discrete Laplace transform as a Taylor series or a Chebyshev series that converges exponentially fast with the truncation order. Each operator of the series is composed of a uniform matrix and two diagonal operators that encode the coefficients of the Laplace transform. These operations are non-unitary, whereas quantum computers are limited to performing unitary quantum gates. One way to overcome this issue is to block-encode the desired operation into a larger unitary operator using ancilla qubits. Each term of the series can be block-encoded, and the summation can be executed using a standard linear combination of unitaries. A second key idea of the protocol is the introduction of an element-wise product of unitaries that makes possible the implementation of the uniform matrix using only a few layers of quantum gates. 

Truncating the series introduces an error $\epsilon>0$ that can be chosen arbitrarily as an input parameter of the problem. The more terms one adds, the larger the quantum circuit becomes, creating a trade-off between accuracy and depth. The depth of a quantum circuit refers to the number of layers of primitive quantum gates.  While depth scaling can indicate the possibility for achieving effective runtime with similar scalings, in practice, it also depends on hardware constraints such as qubit connectivity, the set of implementable primitive quantum gates, the ability to perform quantum gates in parallel and the error correction employed. In this work, we do not account for these constraints, assuming instead that the qubits have full connectivity, are fault-tolerant and that the synthesis of operations is expressed in terms of single- and two-qubit gates without further specification regarding the gate set. 

The presented construction of the QLT requires $O(\log(1/\epsilon))$ controlled diagonal operators and an additional quantum circuit of depth $\tilde{O}(\log(\log(N))\log(1/\epsilon))$\footnote{The $\tilde{O}$ scaling is defined as a $O$ scaling up to a $\text{poly}(\log \log(1/\epsilon))$ term which is described in the proof of Theorem \ref{theorem: discrete Laplace transform} in equation \ref{eq: size,depth}.} and size\footnote{The size is the total number of primitive quantum gates.}  $\tilde{O}(\log(N)\log(1/\epsilon))$ using $O(\log(N)+\log(\log(1/\epsilon)))$ ancilla qubits\footnote{Ancilla qubits are additional qubits that facilitates circuit synthesis.}. In some cases, the circuit depth associated with the diagonal operators scales independently of $N$ as $O(\log(1/\epsilon))$ or as $O(1/\epsilon)$ (see Appendix \ref{sec: QLP with Fourier} or \cite{welch2014efficient,zylberman2024efficient_bis}). In these cases, the circuit depth of the QLT scales with $N$ as  $O(\log \log(N))$. This is doubly-exponentially smaller than the number of arithmetic operations $O(N)$ required by a classical computer to perform the same discrete Laplace transform\footnote{This does not constitute an algorithmic speedup since we are not solving a decision problem. The computational advantage is closer to a "routine speedup", similar to that of the quantum Fourier transform over the classical fast Fourier transform.}. 

In Section \ref{sec: QLP}, the constructions and approximations required to efficiently block-encode the Quantum Laplace transform are presented, and a precise statement of the circuit complexities is provided in Theorem \ref{theorem: discrete Laplace transform}. Section \ref{sec: QLP with complex coefficients} extends the results to the case of discrete Laplace transforms with complex coefficients. Then, the results, along with their applicability, are discussed in Section \ref{sec: discussion}. Additionnally, Appendix \ref{sec: QLP with Fourier} provides an example of diagonal operators that are implementable with $O(\log(1/\epsilon))$ depth, independent of $N$, and Appendix \ref{sec: QLP and continuous} describes how the continuous Laplace transform and its inverse can be approximated using the quantum Laplace transform.

\section{Quantum Laplace Transform}
\label{sec: QLP}

Considering $n$ qubits and two sets of real numbers $(x_1,\hdots,x_N)\in \mathbb{R}^N$, $(y_1,\hdots,y_N)\in \mathbb{R}^N$ with $N=2^n$, we define the Quantum Laplace Transform as the $n$-qubit operation:
\begin{equation}
    \widehat{QLT}= \frac{1}{N}\begin{pmatrix} e^{x_1y_1} &\hdots & e^{x_1y_N} \\ \vdots & \ddots & \vdots \\ e^{x_Ny_1} & \hdots & e^{x_Ny_N} \end{pmatrix}
\label{qdlp}
.\end{equation}
The factor $1/N$ naturally appears in the construction of the operation using primitive quantum gates and as the integration step in the discretization of the continuous Laplace transform (see Appendix \ref{sec: QLP and continuous}).

The $\widehat{QLT}$ is a non-unitary operation and cannot be directly implemented on digital quantum computers. To circumvent this issue, one can block-encode the  $\widehat{QLT}$ operation as part of a larger unitary operation by introducing additional qubits, called ancilla qubits.

\begin{Def}{Block-encoding.}

Let $\hat{A}$ be a $n$-qubit operation, $\alpha,\epsilon \in \mathbb{R}_+$ and $a\in\mathbb{N}$. The $(n+a)$-qubit unitary $\hat{U}$ is called a $(\alpha,a,\epsilon)$-block-encoding of $\hat{A}$, if:
\begin{equation}
    \|\hat{A}-\alpha (\bra{0}^{\otimes a} \otimes \hat{I}_n)\hat{U}(\ket{0}^{\otimes a} \otimes \hat{I}_n)\| \leq \epsilon 
\end{equation}
\end{Def}
The previous definition should be understood as $\hat{U}=\begin{pmatrix}
    \tilde{A}/\alpha&* \\ * &*
\end{pmatrix}$ with $\|\hat{A}-\tilde{A}\|\leq \epsilon$ where the other blocks '$*$' are constructed such that $\hat{U}$ is unitary. This unitary can now be decomposed into a product of primitive quantum gates.

The first idea to construct a block-encoding of the QLT is to approximate each term $e^{x_iy_j}$ using a series expansion of the exponential function. In the following, we consider two types of series, the Taylor series and the Chebyshev series (related to the Jacobi-Anger expansion):
\begin{equation}
\begin{split}
    e^{x_iy_j}&=\sum_{k=0}^{\infty}(x_iy_j)^k/k!
    \\
    e^{x_iy_j}&=e^{\tilde{x}_i\tilde{y}_j}=\sum_{k=0}^\infty (2-\delta_k)I_k(\tilde{x}_i)T_k(\tilde{y}_j)
\end{split}
\end{equation}
where $\delta_k=1$ if $k=0$, $\delta_k=0$ if $k\ge1$ and the coefficients $y_j$ are normalized as $\tilde{y}_j=y_j/y_{\max}\in[-1,1]$ to ensure the existence of the expansion and $\tilde{x}_i=y_{\max}x_i$, where $y_{\max}=\max_{j\in[N]}|y_j|$. $T_k$ is the $k$-th Chebyshev polynomial $T_k(\tilde{y})=\cos(k\arccos(\tilde{y}))$ and $I_k$ is the modified Bessel function of the first kind of order $k$. Two expansions are considered instead of one because the Taylor series is known to be sometimes numerically unstable before reaching its asymptotic regime, even for scalar-valued exponential \cite{moler2003nineteen}.  In order to implement the series, one needs to truncate it, which induces an error that is exponentially small with the truncation order:

\begin{lemma}{(Approximation of the exponential function with truncated series)}
\label{lemma: approx of exp}

Let $x_i,y_j\in\mathbb{R}$, $x_{\max},y_{\max}\in \mathbb{R}_+$ with $|x_i|\leq x_{\max}$, $|y_j|\leq y_{\max}$ and $K\in\mathbb{N}^*$. The Taylor series of order $K$ and the Chebyshev series of order $K$ approximate the exponential function as: \begin{equation}
\begin{split}
    &|e^{x_i y_j}- \sum_{k=0}^K \frac{(x_i y_j)^k}{k!}|\leq e^{x_{\max} y_{\max}} \frac{(x_{\max} y_{\max})^{K+1}}{(K+1)!}
    \\
    &|e^{x_i y_j}- \sum_{k=0}^K (2-\delta_k) I_k(\tilde{x}_i)T_k(\tilde{y}_j)| \leq \frac{4e^{x_{\max}y_{\max}}}{(K+1)!}(\frac{x_{\max} y_{\max}}{2})^{K+1}
\label{Eq: series inequality}
\end{split}
\end{equation}
with $\tilde{x}_i=y_{\max}x_i$ and $\tilde{y}_j=y_j/y_{\max}\in[-1,1]$. For the Chebyshev series, the condition $K+1\ge x_{\max}y_{\max}$ is also required.
\end{lemma}
The proof of this lemma is presented in Appendix \ref{Proof of Lemma approx exp}.

To reach a given accuracy $\epsilon \in ]0,1[$, one can study the function $K(x,\epsilon)$ defined implicitly by $(x/K)^K=\epsilon$ with $x\in\mathbb{R}_+$. Lemma 59 of \cite{gilyen2019quantum} provides bounds on $K(x,\epsilon)$, implying that it is sufficient for the truncation order $K$ of both series to scale as 
\begin{equation}
    K=\Theta(x_{\max}y_{\max}+\frac{\ln(1/\epsilon)}{\ln(e+\ln(1/\epsilon)/{x_{\max}y_{\max}})})
\end{equation}
If $x_{\max}y_{\max}$ is large compared to $\ln(1/\epsilon)$, the scaling is $K=\Theta(x_{\max}y_{\max})$. In the second case, when $x_{\max}y_{\max}$ is small compared to $\ln(1/\epsilon)$, one gets $K=\Theta(\ln(1/\epsilon)/\ln(\ln(1/\epsilon)))$ which provides a more precise expression for $K$ than $\Theta(x_{\max}y_{\max}+\ln(1/\epsilon))$. The parameter $x_{\max}y_{\max}$ is crucial because approximating exponential functions of large arguments is more difficult, especially since block-encodings require normalization. Therefore, we keep track of the parameter $x_{\max}y_{\max}$ in the computations.

After truncation, one can rewrite the QLT operation as a sum of $K$ matrices:
\begin{equation}
\begin{split}
   & \widehat{QLT}\simeq \frac{1}{N} \sum_{k=0}^K \frac{1}{k!} \begin{pmatrix} (x_1y_1)^k &\hdots & (x_1y_N)^k \\ \vdots & \ddots & \vdots \\ (x_Ny_1)^k & \hdots & (x_Ny_N)^k \end{pmatrix} =\sum_{k=0}^K \lambda_k \hat{A}_k. \\ & \text{or} \\
    &\widehat{QLT}\simeq \frac{1}{N} \sum_{k=0}^K (2-\delta_{k}) 
    \begin{pmatrix} I_k(\tilde{x}_1)T_k(\tilde{y}_1) &\hdots & I_k(\tilde{x}_1)T_k(\tilde{y}_N) \\ \vdots & \ddots & \vdots \\ I_k(\tilde{x}_N)T_k(\tilde{y}_1) & \hdots & I_k(\tilde{x}_N)T_k(\tilde{y}_N) \end{pmatrix} =\sum_{k=0}^K \lambda_k' \hat{A}_k'
\end{split}
\end{equation}

with $\lambda_k=(x_{\max}y_{\max})^k/k!$,  $(\hat{A}_k)_{i,j}=(x_iy_j/(x_{\max}y_{\max}))^k/N$, $\lambda_k'=(2-\delta_k)I_{k,\max}$, $(\hat{A}_k')_{i,j}=I_k(\tilde{x}_i)T_k(\tilde{y}_j)/(I_{k,\max}N)$ and $I_{k,\max}=e^{x_{\max}y_{\max}}(x_{\max}y_{\max}/2)^k/k!$ choosen in order to get $|I_k(\tilde{x}_i)T_k(\tilde{y}_j)/I_{k,\max}|\leq 1$ (see Appendix \ref{Proof of Lemma approx exp}).

The sum of $K$ matrices $(\hat{A}_k)_{k\in[K]}$ multiplied by coefficients $\lambda_k\ge0$ can be implemented using a linear combination of unitaries, where each unitary is a block-encoding of the matrix $\hat{A}_k$.
We extend the linear combination of block-encodings introduced in Lemma 52 \cite{gilyen2019quantum} to the case where the block-encodings have different normalizations $\alpha_k>0$ and errors $\epsilon_k>0$.

\begin{lemma}{(Linear combination of block-encodings)}
\label{LCBE}

Let $n,K\in \mathbb{N}^*$, $\lambda_k,\alpha_k,\epsilon_k\in \mathbb{R}_+$ for all $k=0,\hdots,K$, $a\in \mathbb{N}$, $b=\lceil\log_2(K+1)\rceil$ and $\hat{A}=\sum_{k=0}^K \lambda_k \hat{A}_k$ be an $n$-qubit operator. Suppose each $\hat{A}_k$ can be $(\alpha_k,a,\epsilon_k)$-block-encoded in a $(n+a)$-qubit unitary $\hat{U}_k$, and we have access to the $(n+a+b)$-qubits SELECT operation $\hat{S}=\sum_{k=0}^{K}\ket{k}\bra{k}\otimes \hat{U}_k+(\hat{I}_b-\sum_{k=0}^K\ket{k}\bra{k})\otimes \hat{I}_a\otimes I_n$ and to the PREPARE operation $\hat{P}$ acting on $b$ qubits as $\hat{P}\ket{0}^{\otimes b}=\sum_{k=0}^{2^b-1}\mu_k\ket{k}$ with $\mu_k=\sqrt{\lambda_k\alpha_k/\lambda}$ for $k\leq K$,  $\mu_k=0$ for $k>K$ and $\lambda=\sum_{k=0}^{K}\lambda_k\alpha_k$. 

Then, $\hat{W}=(\hat{P}^\dagger\otimes \hat{I}_a\otimes \hat{I}_n)\hat{S}(\hat{P}\otimes \hat{I}_a\otimes \hat{I}_n)$ is a $(\lambda,a+b,\sum_{k=0}^K\lambda_k\epsilon_k )$-block-encoding of $\hat{A}$.
    
\end{lemma}
\begin{proof}

\begin{equation}
\begin{split}
&\|\hat{A}-\lambda(\bra{0}^{\otimes b}\otimes\bra{0}^{\otimes a}\otimes\hat{I}_n)\hat{W}(\ket{0}^{\otimes b}\otimes\ket{0}^{\otimes a}\otimes\hat{I}_n)\|   =\|\hat{A}-\lambda\sum_{k=0}^K\mu_k^2 (\bra{0}^{\otimes a}\otimes\hat{I}_n)\hat{U}_k(\ket{0}^{\otimes a}\otimes\hat{I}_n)\| \\ & = \|\sum_{k=0}^K \lambda_k(\hat{A_k}-\alpha_k(\bra{0}^{\otimes a}\otimes\hat{I}_n)\hat{U}_k(\ket{0}^{\otimes a}\otimes\hat{I}_n))\| \leq \sum_{k=0}^K \lambda_k \epsilon_k
\end{split}
\end{equation}
\end{proof}

To construct the block-encodings of the  $\hat{A}_k$  and $\hat{A}_k'$, one can decompose the matrixes as a product of two non-unitary diagonal operations with the uniform matrix $\hat{A}$, where the entries of $\hat{A}$ are given by $(\hat{A})_{i,j}=1/N$.
\begin{equation}
\begin{split}
    \hat{A}_k&=\begin{pmatrix} (\frac{x_1}{x_{\max}})^k  & & \\  & \ddots &  \\  &  &(\frac{x_N}{x_{\max}})^k \end{pmatrix}\begin{pmatrix} 1/N &\hdots & 1/N \\ \vdots & \ddots & \vdots \\  1/N & \hdots & 1/N \end{pmatrix}\begin{pmatrix} (\frac{y_1}{y_{\max}})^k  & & \\  & \ddots &  \\  &  & (\frac{y_N}{y_{\max}})^k \end{pmatrix} \\ 
     \hat{A}_k'&=\begin{pmatrix} \frac{I_k(\tilde{x}_1)}{I_{k,\max}}  & & \\  & \ddots &  \\  &  &\frac{I_k(\tilde{x}_N)}{I_{k,\max}}\end{pmatrix}\begin{pmatrix} 1/N &\hdots & 1/N \\ \vdots & \ddots & \vdots \\  1/N & \hdots & 1/N \end{pmatrix}\begin{pmatrix}  T_k(\tilde{y}_1)  & & \\  & \ddots &  \\  &  & T_k(\tilde{y}_N) \end{pmatrix}
\end{split}
\end{equation}

In turn, each of these operations can be block-encoded before being multiplied by each other. First, we explain how to block-encode the uniform matrix before discussing the block-encodings of the diagonal operators. 

To block-encode the uniform matrix $(\hat{A})_{i,j}=1/N$, we define the element-wise product of two $n$-qubit unitaries $\hat{U}$ and $\hat{V}$ as the matrix $\hat{U}\odot\hat{V}$, whose entries are given by $(\hat{U}\odot\hat{V})_{i,j}=(\hat{U})_{i,j}(\hat{V})_{i,j}$. The following lemma introduces an efficient method for implementing the element-wise product of unitaries:

\begin{lemma}{(Element-wise product of unitaries)}

Let $n\in\mathbb{N}^*$, $\hat{U}$, $\hat{V}$ be two unitaries acting on an $n$-qubit register $\{q_i\}_{i\in[n]}$. Consider an additional register of $n$ ancilla qubits $\{a_i\}_{i\in[n]}$ and the operation $\hat{C}=\bigotimes_{i=1}^n(CNOT_{q_i\rightarrow a_i})$ which is a layer of parallel CNOT gates. The $CNOT_{q_i\rightarrow a_i}$ denotes the $CNOT$ gate controlled by the $i$-th qubit $q_i$ and applied to the $i-$th ancilla qubit $a_i$. 

The unitary  $\hat{W}=\hat{C}(\hat{U}\otimes\hat{V})\hat{C}$  is a $(1,n,0)-$block-encoding of $\hat{U}\odot\hat{V}$.
\label{lemma:element-wise product}
\end{lemma}
\begin{proof}

Remark that for $\ket{x}=\bigotimes_{i=0}^n\ket{x_i}$ with $x_i\in\{0,1\},$ $\hat{C}\ket{x}\otimes\ket{0}^{\otimes n}=\ket{x}\otimes\ket{x}$, then:
    
\begin{equation}
\begin{split}
    (\bra{0}^{\otimes n} \otimes \bra{x})\hat{W}(\ket{0}^{\otimes n} \otimes \ket{x'})&=(\bra{0}^{\otimes n} \otimes \bra{x})\hat{C}(\hat{U}\otimes\hat{V})\hat{C}(\ket{0}^{\otimes n} \otimes \ket{x'}) \\
    &=(\bra{x} \otimes \bra{x})(\hat{U}\otimes\hat{V})(\ket{x'} \otimes \ket{x'}) \\&=\bra{x} \hat{U} \ket{x'} \bra{x} \hat{V} \ket{x'} = \hat{U}_{x,x'}\hat{V}_{x,x'}= (\hat{U}\odot\hat{V})_{x,x'}
\end{split}
\end{equation}
\end{proof}

This lemma provides a direct way to implement the uniform matrix by noting that $(\hat{A})_{i,j}=1/N$ is the element-wise product of two Hadamard towers $(\hat{H}^{\otimes n}\odot \hat{H}^{\otimes n})_{i,j}=( \hat{H}^{\otimes n})_{i,j}( \hat{H}^{\otimes n})_{i,j}=1/N$, with $\hat{H}=\frac{1}{\sqrt{2}}\begin{pmatrix}
    1&1\\1&-1
\end{pmatrix}$.  Therefore, one can implement a $(1,n,0)$-block-encoding of $(\hat{A})_{i,j}=1/N$ with $2n$ CNOT gates, $2n$ Hadamard gates, and a quantum circuit depth equal to $3$.

As discussed in the following section \ref{sec: discussion}, block-encoding a non-unitary diagonal operator can be done by many different methods, with complexities depending directly on the properties of the eigenvalues (sparsity, smoothness,...) \cite{zylberman2024efficient_bis}. In some cases, the depth is independent of the number of qubits and scales only with the implementation error $\epsilon$, or logarithmically with the number of qubits $n$ (see Appendix \ref{sec: QLP with Fourier} or \cite{zylberman2024efficient_bis}). To keep the results general enough, we assume that we have access to the block-encodings $ \hat{U}_{X,k}, \hat{U}_{Y,k}, \hat{U}_{I,k}, \hat{U}_{T,k}$ of the non-unitary diagonal operators:
\begin{equation}
\begin{split}
   & \hat{X}_k=\sum_{i=0}^{N-1}(\frac{x_i}{x_{\max}})^k \ket{i}\bra{i}  \\ &\hat{Y}_k=\sum_{j=0}^{N-1}(\frac{y_j}{x_{\max}})^k \ket{j}\bra{j} \\
    &\hat{I}_k=\sum_{i=0}^{N-1}\frac{I_k(\tilde{x}_i)}{I_{k,\max}}\ket{i}\bra{i},  \\ & \hat{T}_k=\sum_{j=0}^{N-1}T_k(\tilde{y_j})\ket{j}\bra{j}
\label{diag_ope}
\end{split}
\end{equation}
These diagonal operators have eigenvalues bounded by $1$, implying that no additional normalization constant is necessary to block-encode them. Consequently, we suppose that the unitaries $ \hat{U}_{X,k}$, $\hat{U}_{I,k}$ are  $(1,a_{be},\epsilon_{k,1})$-block-encodings of $\hat{X}_k$, $\hat{I}_k$ and $\hat{U}_{Y,k},\hat{U}_{T,k}$ are $(1,a_{be},\epsilon_{k,2})$-block encodings of $\hat{Y}_k$ and $\hat{T}_k$, where $\epsilon_{k,1},\epsilon_{k,2}\in\mathbb{R}_+$ and $a_{be}\in\mathbb{N^*}$. For the implementation of an arbitrary discrete Laplace transform, we estimate the complexity in terms of the number of times these operators needs to be implemented.

Now, using the product of the block-encodings of the non-unitary diagonal operators with the uniform matrix and the following lemma, one can construct $(1,n+2a_{be},\epsilon_{k,1}+\epsilon_{k,2})$-block-encoding of $\hat{A}_k$ and $\hat{A}_k'$. We recall Lemma 53 of \cite{gilyen2019quantum}:

\begin{lemma} (Product of block-encoded matrices)
\label{lemma: product of be matrices}
If $\hat{U}$ is a $(\alpha,a,\epsilon_1)$-block-encoding of an $n$-qubit operator $\hat{A}$ and $\hat{V}$ is a $(\beta,b,\epsilon_2)$-block-encoding of a $n$-qubits operator $\hat{B}$ then $(\hat{I}_a\otimes\hat{U})(\hat{I}_b\otimes \hat{V})$ is a $(\alpha\beta,a+b,\alpha\epsilon_2+\beta\epsilon_1)$-block-encoding of $\hat{A}\hat{B}$.
\end{lemma}

\begin{figure}
    \centering
\includegraphics[scale=0.65]{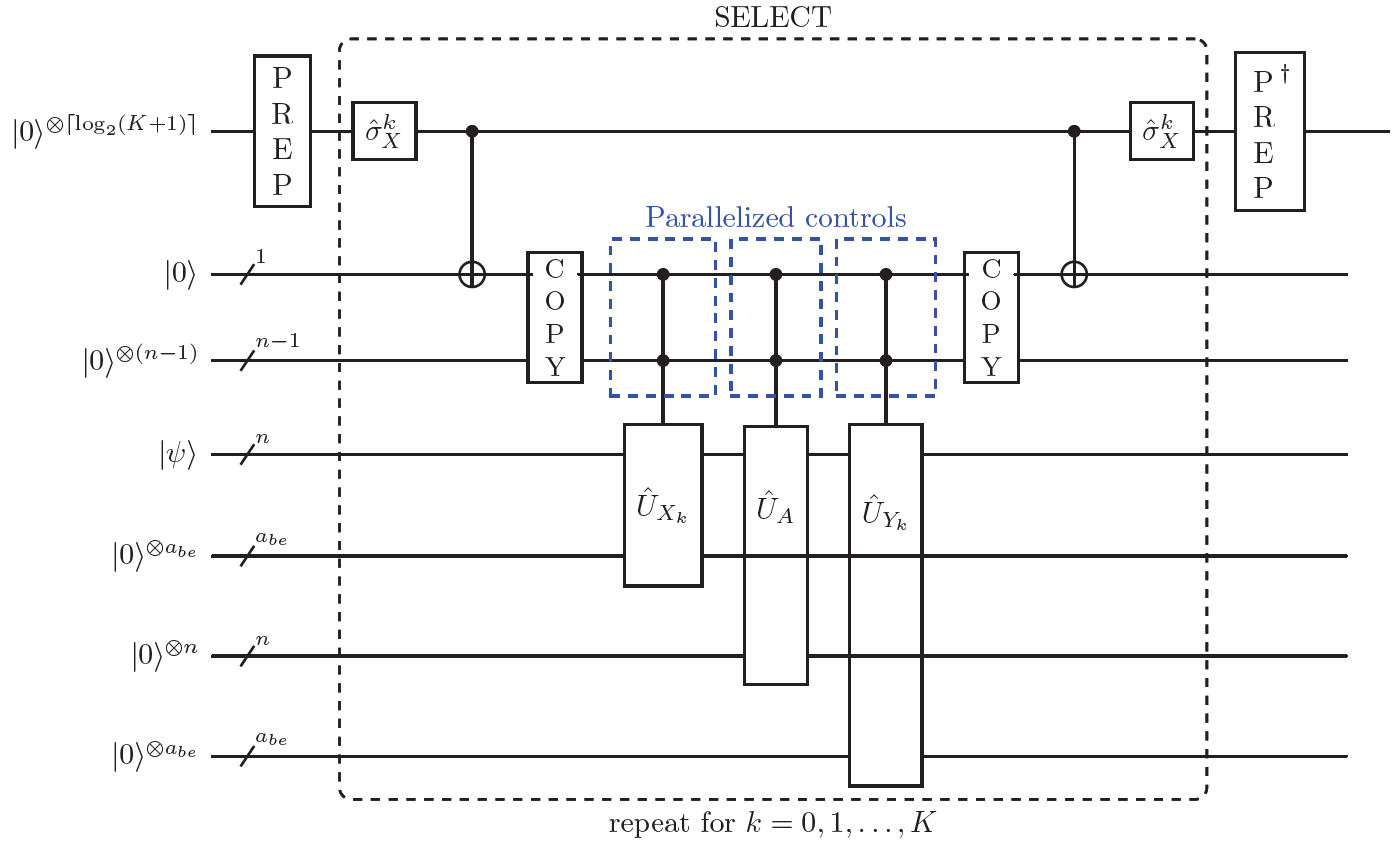}
	\caption{Scheme of the quantum circuit associated with the block-encoding of the Quantum Laplace Transform. The target state, denoted by $\ket{\psi}$, to which the QLT is applied is an n-qubit state. The other qubit registers are zeroed ancilla register that enable the block-encoding of the QLT. The operator $\hat{U}_{X_k}$ and $\hat{U}_{Y_k}$ represent the $(1,a_{be},\epsilon_{k,1,2})$-block-encodings of the diagonal operators $\hat{X}_k$ and $\hat{Y}_k$ defined in Eq.\ref{diag_ope}. They can be replaced by $\hat{U}_{I_k}$ and $\hat{U}_{T_k}$ for the Chebyshev expansion case. $\hat{U}_A$ is the $(1,n,0)$-block-encoding of the uniform matrix $\hat{A}$ with entries $\hat{A}_{i,j}=1/2^n$. The parallelized controls maintain the depth scaling of the block-encodings: $n-1$-zeroed ancilla, which are "copies" of the control ancilla, enables to control several gates in parallel (see Appendix \ref{sec:parallel control}). The COPY operator consists of $n-1$ CNOT gates arranged in $O(\log(n))$ layers (see the operator $\hat{C}$ in lemma \ref{copy lemma}). The operator $\sigma_X^k$ is a tensor product of $X$-Pauli gates that transforms the anti-control of the SELECT operation into a control: $\sigma_X^k=\bigotimes_{i=0}^{\lceil \log_2(K+1) \rceil-1}\hat{\sigma}_X^{1-k_i}$, where $k=\sum_{i=0}^{\lceil \log_2(K+1) \rceil-1}k_i2^i$, $k_i\in\{0,1\}$. The PREP operator is a quantum state preparation that allows performing the weighted sum of the block-encodings, as defined in the Linear Combination of block-encodings lemma \ref{LCBE}.}
	\label{fig:qc for QLT}
\end{figure}

Finally, the linear combination of the block-encodings enables us to approximate the QLT. The following theorem summarizes this result and the computational resources required, while Figure \ref{fig:qc for QLT} provides a schematic view of the quantum circuit implementing the QLT.

\begin{theorem}{Quantum Laplace Transform}
\label{theorem: discrete Laplace transform}

Let $n\in\mathbb{N}^*$, $a_{be} \in \mathbb{N}$, $\epsilon\in \mathbb{R}_+$, $(x_1,\hdots,x_N)\in \mathbb{R}^N$, $(y_1,\hdots,y_N)\in \mathbb{R}^N$ where $N=2^n$, $x_{\max}=\max_{i\in [N]} |x_i|$, $y_{\max}=\max_{i\in [N]} |y_i|$, $a_{be}=O(n)$ 

\begin{itemize}
    \item Taylor expansion: the $n$-qubit $\widehat{QLT}$ defined by Eq.(\ref{qdlp}) can be $(e^{x_{\max}y_{\max}},2n+2a_{be}+\lceil\log_2(K+1)\rceil,\epsilon)$-block-encoded using $O(K)$ one-controlled operators $C(\hat{U}_{X,k})$ and  $C(\hat{U}_{Y,k})$, where $\hat{U}_{X,k}$, $\hat{U}_{Y,k}$ are the $(1,a_{be},\epsilon/(6e^{x_{\max}y_{max}}))$-block-encodings of $\hat{X}_k$ and $\hat{Y}_k$  defined by Eq.\ref{diag_ope}, and an additional quantum circuit of size $\tilde{O}(nK) $ and depth $\tilde{O}(\log(n)K)$ with $K=\Theta(x_{\max}y_{\max}+\frac{\ln(1/\epsilon)}{\ln(e+\frac{\ln(1/\epsilon)}{x_{\max}y_{\max}})})$.

\item Chebytshev expansion: the $n$-qubit $\widehat{QLT}$ defined by Eq.(\ref{qdlp}) can be $(e^{x_{\max}y_{\max}}(2e^{x_{\max}y_{\max}/2}-1),2n+2a_{be}+\lceil\log_2(K+1)\rceil,\epsilon)$-block-encoded using $O(K)$ one-controlled operators $C(\hat{U}_{I,k})$ and  $C(\hat{U}_{T,k})$, where $\hat{U}_{I,k}$, $\hat{U}_{T,k}$ are the $(1,a_{be},\epsilon/(6e^{x_{\max}y_{\max}}(2e^{x_{\max}y_{\max}/2}-1)))$-block-encodings of $\hat{I}_k$ and $\hat{T}_k$ defined by Eq.\ref{diag_ope}, and an additional quantum circuit of size $\tilde{O}(nK) $ and depth $\tilde{O}(\log(n)K)$ with $K=\Theta(x_{\max}y_{\max}+\frac{\ln(1/\epsilon)}{\ln(e+\frac{\ln(1/\epsilon)}{x_{\max}y_{\max}})})$.

The notation $\tilde{O}$ corresponds to a $O$ scaling up to a factor polylogarithmic in $K$.

\end{itemize}
\end{theorem}

\begin{proof}
    Consider the approximation of the QLT by a truncated Taylor series (the computations are similar for the truncated Chebytshev series). As explained in the previous paragraphs, the uniform matrix can be $(1,n,0)$-block-encoded. We assume that each diagonal operator $\hat{X}_k$ can be $(1,a_{be},\epsilon_{k,1})$-block-encoded and each $\hat{Y}_k$ can be  $(1,a_{be},\epsilon_{k,2})$-block-encoded where $\epsilon_{k,1},\epsilon_{k,2}\in \mathbb{R}_+$ are controllable error parameters for all $k=0,...,N$.  The product of block-encodings (lemma \ref{lemma: product of be matrices}) implies that each $\hat{A}_k$ can be $(1,n+2a_{be},\epsilon_{k,1}+\epsilon_{k,2})$-block-encoded. The linear combination of block-encodings (lemma \ref{LCBE}) produces $\hat{W}$ a $(\lambda,n+2a_{be}+\lceil\log_2(K+1)\rceil,\sum_{k=0}^K\lambda_k(\epsilon_{k,1}+\epsilon_{k,2}))$-block-encoding of the operator $\sum_{k=0}^K \lambda_kA_k$ with $\lambda_k=(x_{\max}y_{\max})^k/k!$, $\lambda=\sum_{k=0}^K\lambda_k$.
In the following, we prove that $\hat{W}$ is a $(e^{x_{\max}y_{\max}},b,\epsilon)$-block-encoding of the QLT operation with  $b=n+2a_{be}+\lceil\log_2(K+1)\rceil$.

First, notice that the QLT operation is approximated by a truncated Taylor series $\sum_{k=0}^K\lambda_k \hat{A}_k$ which is, in turn, approximately block-encoded via $\hat{W}$. We use the triangle inequality to express the two differences as follows:
\begin{equation}
\begin{split}
    &\|\widehat{QLT}-\lambda(\bra{0}^{\otimes b}\otimes \hat{I}_n)\hat{W}(\ket{0}^{\otimes b}\otimes \hat{I}_n)\| \\
    &\leq \|\widehat{QLT}-\sum_{k=0}^K\lambda_k\hat{A}_k\|+\|\sum_{k=0}^K\lambda_k\hat{A}_k-\lambda(\bra{0}^{\otimes b}\otimes \hat{I}_n)\hat{W}(\ket{0}^{\otimes b}\otimes \hat{I}_n)\|
\label{eq:DQLP proof}
\end{split}
\end{equation}
 The first term of this inequality can be bounded using the norm inequality $\|\hat{M}\| \leq \|\hat{M}\|_F\leq N \max_{i,j=0,...,N-1}|m_{i,j}|$ where $\|.\|$ is the spectral norm and $\|.\|_F$ is the Frobenius norm.
\begin{equation}
     \|\widehat{QLT}-\sum_{k=0}^K\lambda_k\hat{A}_k\|\leq \max_{i,j=1,\hdots,N} |e^{x_iy_j}-\sum_{k=0}^K\frac{(x_iy_j)^k}{k!}|\leq e^{x_{\max}y_{\max}}\frac{(x_{\max }y_{\max})^{K+1}}{(K+1)!}\leq \epsilon/3
\label{Eq exp -Tk}
\end{equation}
for $K=\Theta(x_{\max}y_{\max}+\frac{\ln(1/\epsilon)}{\ln(e+\frac{\ln(1/\epsilon)}{x_{\max}y_{\max}})})$\footnote{the factor $3$ in the expression $\log(3/\epsilon)$ is asymptotically negligible.}\cite{gilyen2019quantum}. By definition of the block-encoding, the second term of inequality \ref{eq:DQLP proof} is bounded by:
\begin{equation}
    \sum_{k=0}^K \lambda_k(\epsilon_{k,1}+\epsilon_{k,2})\leq 2e^{x_{\max}y_{\max}}\max_{k\in [K]}(\epsilon_{k,1},\epsilon_{k,2})\leq \epsilon/3
\end{equation}
where $\epsilon_{k,1},\epsilon_{k,2}=\epsilon/(6e^{x_{\max}y_{\max}})$ for all $k\in[K]$.  It follows that $\hat{W}$ is a $(\lambda,b,2\epsilon/3)$-block-encoding of the QLT operation. Next, observe the following property of block-encodings: If $\hat{U}$ is a $(\alpha,a,\epsilon_1)$-block-encoding of a matrix $\hat{A}$ and $|\alpha'-\alpha|\leq\epsilon_2$ with $\alpha'>\|A\|$, then $\hat{U}$ is a $(\alpha',a,\epsilon_1+\epsilon_2)$-block encoding of $\hat{A}$:
\begin{equation}
\begin{split}
   & \|\hat{A}-\alpha'(\bra{0}^{\otimes a}\otimes \hat{I}_n)\hat{U}(\ket{0}^{\otimes a}\otimes \hat{I}_n)\| \\&\leq \|\hat{A}-\alpha(\bra{0}^{\otimes a}\otimes \hat{I}_n)\hat{U}(\ket{0}^{\otimes a}\otimes \hat{I}_n)\|+|\alpha-\alpha'| \|(\bra{0}^{\otimes a}\otimes \hat{I}_n)\hat{U}(\ket{0}^{\otimes a}\otimes \hat{I}_n) \| \\& \leq \epsilon_1+\epsilon_2
\end{split}
\end{equation}
where $\|(\bra{0}^{\otimes a}\otimes \hat{I}_n)\hat{U}(\ket{0}^{\otimes a}\otimes \hat{I}_n) \| \leq 1$ since $\hat{U}$ is unitary.

Thus, from Eq.\ref{Eq exp -Tk}, $|\lambda-e^{x_{\max}y_{\max}}|\leq \epsilon/3$ implies that $\hat{W}$ is a $(e^{x_{\max}y_{\max}},b,\epsilon)$-block-encoding of the QLT operation.

We shift our focus to the circuit complexity associated with the various operations required for the block-encoding $\hat{W}$. Specifically, we demonstrate how the introduction of $n$ additional ancilla qubits enables the implementation of the block-encoding $\hat{W}$ using $O(K)$ controlled-block-encodings of $\hat{X}_k$ and $\hat{Y}_k$ and a quantum circuit of size $\tilde{O}(Kn)$ and depth $\tilde{O}(K\log(n))$, where $\tilde{O}$ correspond to a $O$ scaling up to a polylogarithmic factor in $K$, which is explicitely determined in the following. 

The block-encoding $\hat{W}$ of the QLT is constructed from the linear combination of the $K+1$ block-encodings $\hat{U}_k$ of the operators $\hat{A}_k$. The linear combination of block-encodings consists of two PREPARE routines acting on $\lceil\log_2(K+1)\rceil$ qubits and a SELECT$(\hat{U}_0,\hdots,\hat{U}_K)$ routine. The PREPARE routines can be implemented exactly with a circuit size $O(K)$ and circuit depth $O(K)$ without requiring additional ancilla qubits \cite{mottonen2004transformation}. The SELECT routine requires implementing a number $K$ of $\lceil\log_2(K+1)\rceil$-controlled-$\hat{U}_k$ operations \cite{childs2012hamiltonian}:
\begin{equation}
\begin{split}
&SELECT=\sum_{k=0}^{ K}\ket{k}\bra{k}\otimes\hat{U}_k=\prod_{k=0}^K \Biggl( \ket{k}\bra{k}\otimes \hat{U}_k+(\hat{I}-\ket{k}\bra{k})\otimes \hat{I} \Biggl) \\&
=\prod_{q_1=0}^1\prod_{q_2=0}^1 \hdots \prod_{q_{\lceil log_2(K+1)\rceil}=0}^1 \Biggl( ( \bigotimes_{i=1}^{\lceil log_2(K+1)\rceil}\ket{q_i}\bra{q_i})\otimes \hat{U}_k+ (\hat{I}-\bigotimes_{i=1}^{ \lceil log_2(K+1)\rceil}\ket{q_i}\bra{q_i})\otimes \hat{I} \Biggl)\\
&=\prod_{q_1=0}^1\prod_{q_2=0}^1 \hdots \prod_{q_{\lceil log_2(K+1)\rceil}=0}^1 \Biggl( \bigotimes_{i=1}^{\lceil log_2(K+1)\rceil}  \hat{\sigma}_{X}^{1-q_i}\otimes \hat{I} \Biggl) \Biggl( (\bigotimes_{i=1}^{\lceil log_2(K+1)\rceil}\ket{q_i=1}\bra{q_i=1})\otimes \hat{U}_k \\&+(\hat{I}-(\bigotimes_{i=1}^{ \lceil log_2(K+1)\rceil}\ket{q_i=1}\bra{q_i=1}))\otimes \hat{I} \Biggl) \Biggl(\bigotimes_{i=1}^{\lceil log_2(K+1)\rceil}\hat{\sigma}_{X}^{1-q_i} \otimes \hat{I} \Biggl) \\
&=\prod_{k=0}^K (\hat{\sigma}_{X}^{k}\otimes \hat{I}) C_{q_1\hdots q_{\lceil log_2(K+1)\rceil}}(\hat{U}_k)(\hat{\sigma}_{X}^{k}\otimes \hat{I})
\end{split}
\end{equation}
where $\hat{\sigma}_{X}^{k}=\bigotimes_{i=1}^{\lceil log_2(K+1)\rceil}\hat{\sigma}_{X}^{1-q_i}$ is a one layer operator, $k=\sum_{i=1}^{\lceil \log_2(K+1) \rceil } q_i 2^{i-1}$ and $C_{q_1\hdots q_{\lceil log_2(K+1)\rceil}}(\hat{U}_k)$ is the control $\hat{U}_k$ operation that applies $\hat{U}_k$ only if all the qubits $q_i$ are in state $\ket{1}$. 

The multi-controlled-$\hat{U}_k$ operation can be transformed into two multi-controlled NOT gates and a single-controlled-$\hat{U}_k$ using one ancilla qubit $q_A$  (Lemma 7.11 in \cite{barenco1995elementary}).
\begin{equation}
\begin{split}
&C_{q_1\hdots q_{\lceil log_2(K+1)\rceil}}(\hat{U}_k)\otimes \hat{I}_{q_A} \\ &=\Biggl(C_{q_1\hdots q_{\lceil log_2(K+1)\rceil}}(\hat{\sigma}_{X,q_A})\otimes \hat{I}_n\Biggl)\Biggl( \hat{I}_{\lceil \log_2(K+1) \rceil }\otimes C_{q_A}(\hat{U}_k) \Biggl) \Biggl(C_{q_1\hdots q_{\lceil log_2(K+1)\rceil}}(\hat{\sigma}_{X,q_A}) \otimes \hat{I}_n\Biggl) 
\end{split}
\end{equation}
where $\hat{\sigma}_{X,q_A}$ is a $X$-Pauli gates and $\hat{I}_{q_A}$ is the identity operator, both acting on the ancilla qubit $q_A$.

Thus, the $K$ multi-controlled $\hat{U}_k$ are decomposed into $2K$ multi-controlled NOT gates and $K$ single-controlled $\hat{U}_k$.  Depending on the number of available ancilla, the multi-controlled NOT gates $C_{q_1\hdots q_{\lceil log_2(K+1)\rceil}}(\hat{\sigma}_X)$ can be implemented with different methods. For a $\lceil\log_2(K)\rceil$-controlled NOT gate, most of the quantum circuits have a size in $O(\log(K)\text{polylog}(\log(K)))$ and a depth in $O(\text{polylog}(\log(K)))$ using no additional ancilla qubits \cite{claudon2024polylogarithmic,nie2024quantum}, one additional ancilla qubits \cite{claudon2024polylogarithmic,nie2024quantum}, two additional ancilla qubits \cite{khattar2024rise} or $n$ additional ancilla qubits \cite{he2017decompositions}. Therefore, the $2K$ $\lceil\log_2(K)\rceil$-controlled NOT gates require $O(K\log(K)\text{polylog}(\log(K)))$ quantum gates and a circuit depth  $O(K\text{polylog}(\log(K)))$.  The zeroed ancilla qubits used for the implementation of the multi-controlled-NOT operations can be taken directly from the one that are introduced to parallelize the control as explained in the next paragraph and as shown in Figure \ref{fig:qc for QLT}.

Since $\hat{U}_k$ is equal to the product of three block-encodings, the implementation of the one-controlled-$\hat{U}_k$ is directly obtained by multiplying the single-controlled-$\hat{U}_{X,k}$, the single-controlled block-encoding of the uniform matrix $\hat{U}_A$ and the single-controlled-$\hat{U}_{Y,k}$. 
Note that a single qubit cannot control multiple gates simultaneously. To overcome this issue, we introduce ancilla qubits that help to  parallelize the control operation. This procedure is standard and detailed in Appendix \ref{sec:parallel control}. To summarize, if $\hat{U}$ is a $n$-qubit unitary implementable with a quantum circuit of size $O(s(n))$ and depth $O(d(n))$ then, the one-control-$\hat{U}$ operation $C_{a_1}(U)=\hat{I}_n\otimes\ket{a_1=0}\bra{a_1=0}+\hat{U}\otimes\ket{a_1=1}\bra{a_1=1}$ is implementable with a quantum circuit of size $O(s(n)+n)$ and depth $O(d(n)+ \log(n))$ by using $n-1$ additional ancilla qubit. This protocol uses two COPY operators, noted $\hat{C}$ in Appendix \ref{sec:parallel control}, which are composed of $2n$ CNOT gates implemented in logarithmic depth $O(\log(n))$.

As explained above, the block-encoding of $\hat{A}$ can be constructed using $n$ ancilla qubit and the element-wise product of two Hadamard tower $\hat{U}_A=\hat{C}\hat{H}^{\otimes 2n}\hat{C}$, where $\hat{C}=\bigotimes_{i=1}^n(CNOT_{q_i\rightarrow a_i})$. This block-encoding is made of $4n$ primitive quantum gates arranged in three layers. Therefore, the parallelization-of-control procedures implies that the one-controlled $\hat{U}_A$ operator is implementable with a size scaling as $O(n)$ and a depth as $O(\log(n))$.

Then, we need to implement the two one-controlled-$(1,a_{be},\epsilon/(6e^{x_{\max}y_{\max}}))$-block-encodings $\hat{U}_{X,k}$ and $\hat{U}_{Y,k}$. We note that a similar parallelization of controls can be employed to implement $C(\hat{U}_{X,k})$ and  $C(\hat{U}_{Y,k})$. The condition $a_{be}=O(n)$ implies that the $n-1$ qubits already introduced for $C(\hat{U}_A)$ are sufficient to preserve the depth scaling of $\hat{U}_{X,k}$ and $\hat{U}_{Y,k}$,  up to a constant factor $\lceil n_a/n\rceil=O(1)$ (see Lemma \ref{lemma : parallel control of one unitary}).

Summarizing, the overall circuit is made of two PREPARE operations,  $2K$ $\lceil\log_2(K+1)\rceil-$-controlled NOT gates, $2K$ copy operations, $2K$ single-control diagonal operations and $K$ single-control block-encoding to the uniform matrice. Thus, one needs $O(K)$ single-control diagonal operations and a quantum circuit of size and depth:
\begin{equation}
\begin{split}
\label{eq: size,depth}
s(n,\epsilon)&=O \Biggl(K+K\log(K)\text{polylog}(\log(K))+Kn+Kn \Biggl)\\ &=O \Biggl(K(n+\log(K)\text{polylog}(\log(K))) \Biggl)=\tilde{O}(Kn)
\\ d(n,\epsilon)&=O \Biggl(K+K\text{polylog}(\log(K))+K\log(n)+K \Biggl) \\
&=O \Biggl(K(\log(n)+\text{polylog}(\log(K))) \Biggl)=\tilde{O}(K\log(n))
\end{split}
\end{equation}
 If we note $s_U(n,\epsilon)$ and $d_U(n,\epsilon)$, the maximum size and depth of the block-encodings $\hat{U}_{X,k}$,$\hat{U}_{Y,k}$ then, the overall quantum circuit size is $O(s(n,\epsilon)+Ks_U(n,\epsilon/(6e^{x_{\max}y_{\max}})))$ and the depth is $O(d(n,\epsilon)+ Kd_U(n,\epsilon/(6e^{x_{\max}y_{\max}})))$. 

The proof is similar for the Chebytchev expansion.
\end{proof}

\begin{remark}
    As explained in the previous proof, the condition on the ancilla $a_{be}=O(n)$ enables the implementation of the $C(\hat{U}_{X,k})$ and $C(\hat{U}_{Y,k})$ (resp. $C(\hat{U}_{I,k})$ and $C(\hat{U}_{T,k})$) with the same depth scalings as $\hat{U}_{X,k}$ and $\hat{U}_{Y,k}$ (resp. $\hat{U}_{I,k}$  and $\hat{U}_{T,k}$)  up to an $O(\log(n))$ term by using the parallelization of control techniques presented in Appendix \ref{sec:parallel control}.
\end{remark}

Consequently, when the block-encodings of the non-unitary diagonal operators are implementable with a polylogarithmic depth in $n$ and $1/\epsilon$, the QLT is also implementable in $\text{polylog}(n,1/\epsilon)$ depth on a quantum computer. In some cases, the diagonal operators can be implemented with a circuit depth independent of $N$ (see Appendix \ref{sec: QLP with Fourier} or \cite{welch2014efficient,zylberman2024efficient_bis}). This implies that the QLT is implementable with a depth scaling with $N$ as $O(\log(\log(N)))$. The scheme of the quantum circuit associated to the block-encoded of the QLT is presented in Figure \ref{fig:qc for QLT}. In the following section \ref{sec: discussion}, we generalize these results to QLT with complex coefficients.

\section{Quantum Laplace Transform with complex coefficients}
\label{sec: QLP with complex coefficients}

For the Taylor expansion, the implementation of the QLT with complex coefficients is almost identical as in the real case. The only difference appears in the non-unitary diagonal operators defined in Eq.(\ref{diag_ope}), which have complex eigenvalues instead of real ones. To adapt the Chebytshev expansion of the exponential for the complex case, the coefficients need to be redefined, because the Chebyshev polynomials $T_k$ are defined on $[-1,1]$. Hence, the real and imaginary parts of $y_j=\Re(y_j)+\mathbf{i}\Im(y_j)$, with $\mathbf{i}^2=-1$, must be introduced:
\begin{equation}  
e^{x_iy_j}=e^{x_i\Re(y_j)}e^{\mathbf{i}  x_i\Im(y_j)}=e^{\tilde{x}_i\tilde{y}_j}e^{\tilde{x_i}'\tilde{y_j}' }
\end{equation}
with the new variables:
\begin{equation}
\begin{split}
    \tilde{x}_i&=x_i\max_{j\in[N]}|\Re(y_j)|\in \mathbb{C} \\
    \tilde{y}_j&=\Re(y_j)/\max_{j\in[N]}|\Re(y_j)| \in [-1,1] \\
    \tilde{x}_i'&=\mathbf{i}x_i\max_{j\in[N]}|\Im(y_j)|\in \mathbb{C} \\
    \tilde{y}_j'&=\Im(y_j)/\max_{j\in[N]}|\Im(y_j)| \in [-1,1]
\end{split}
\end{equation}
Thanks to this change of variable, each of the two exponential can be expanded in a Chebytshev series $e^{\tilde{x}_i\tilde{y}_j}=\sum_{k=0}^{\infty}(2-\delta_k)I_{k}(\tilde{x}_i)T_k(\tilde{y}_j)$ and $e^{\tilde{x_i}'\tilde{y_j}'}=\sum_{k'=0}^{\infty}(2-\delta_{k'})I_{k'}(\tilde{x}_i')T_{k'}(\tilde{y}_j)'$. The error introduced by truncating these series can be bounded by using the relationship between the modified Bessel function of the first kind $I_k$ and the Bessel function of first kind $J_k$: for all $k\in \mathbb{N}$ and $z\in \mathbb{C}$ $I_k(z)=(-\mathbf{i})^kJ_k(\mathbf{i}z)$   (Eq. 9.3.6 in \cite{abramowitz1968handbook}). Then, the Bessel function is bounded as $|J_k(z)|\leq |z/2|^ke^{|\Im(z)|}/k!$ for all $k\in\mathbb{N}^*$ and $z\in\mathbb{C}$ (Eq. 9.1.62 in \cite{abramowitz1968handbook}). Using these results, one can show for all $1\leq i,j \leq N$:

\begin{equation}
\label{inequality 2Dcase}
\begin{split}
    &|e^{x_iy_j}-\sum_{k=0}^K\sum_{k'=0}^{K}(2-\delta_k)(2-\delta_{k'})I_{k}(\tilde{x}_i)I_{k'}(\tilde{x}_i')T_k(\tilde{y}_j)T_{k'}(\tilde{y}_j') \leq \frac{8 e^{\frac{5}{2}x_{\max}y_{\max}}}{(K+1)!}|\frac{x_{\max}y_{\max}}{2}|^{K+1}
\end{split}
\end{equation}
which is asymptotically $\epsilon$-small for $K=\Theta(x_{\max}y_{\max}+\frac{\ln(1/\epsilon)}{\ln(e+\frac{\ln(1/\epsilon)}{x_{\max}y_{\max}})})$ (see Appendix \ref{proof: 2Dcase} for the proof of this inequality). The double sum induces a quadratic overhead in the number of diagonal operators compared to the QLT with real coefficients. These results are summarized in the following theorem:

\begin{theorem}{Quantum Laplace Transform with complex coefficients}
\label{theorem: discrete Laplace transform with complex coefficients}

Let $n\in \mathbb{N}$, $a_{be}\in \mathbb{N}$, $\epsilon\in \mathbb{R}_+$, $(x_1,\hdots,x_N)\in \mathbb{C}^N$, $(y_1,\hdots,y_N)\in \mathbb{C}^N$ with $N=2^n$, $x_{\max}=\max_{i\in [N]} |x_i|$, $y_{\max}=\max_{i\in [N]} |y_i|$, $a_{be}=O(n)$.
\begin{itemize}
    \item Taylor expansion: the $n$-qubit $\widehat{QLT}$ with complex coefficients $\{x_i\}_{i\in[N]},\{y_j\}_{j\in[N]}$ can be $(e^{x_{\max}y_{\max}},2n+2a_{be}+\lceil\log_2(K+1)\rceil,\epsilon)$-block-encoded using $O(K)$ one-controlled operators $C(\hat{U}_{X,k})$ and  $C(\hat{U}_{Y,k})$, where $\hat{U}_{X,k}$, $\hat{U}_{Y,k}$ are the $(1,a_{be},\epsilon/(6e^{x_{\max}y_{\max}}))$-block-encodings of $\hat{X}_k=\sum_{i=1}^{N}(x_i/x_{\max})^k\ket{i}\bra{i}$ and $\hat{Y}_k=\sum_{j=1}^{N}(y_j/y_{\max})^k\ket{j}\bra{j}$, and an additional quantum circuit of size $\tilde{O}(nK)$ and depth $\tilde{O}(\log(n)K)$ with $K=\Theta(x_{\max}y_{\max}+\frac{\ln(1/\epsilon)}{\ln(e+\frac{\ln(1/\epsilon)}{x_{\max}y_{\max}})})$.

\item Chebytchev expansion: the $n$-qubit $\widehat{QLT}$ with complex coefficients $\{x_i\}_{i\in[N]},\{y_j\}_{j\in[N]}$ can be $(\alpha,2n+2a_{be}+\lceil2\log_2(K+1)\rceil,\epsilon)$-block-encoded using $O(K^2)$ one-controlled operators $C(\hat{U}_{I,k,k'})$ and  $C(\hat{U}_{T,k,k'})$, where $\hat{U}_{I,k,k'}$, $\hat{U}_{T,k,k'}$ are the $(1,a_{be},\epsilon/(6\alpha))$-block-encodings of $\hat{I}_{k,k'}$ and $\hat{T}_{k,k'}$ and an additional quantum circuit of size $\tilde{O}(nK^2) $ and depth $\tilde{O}(\log(n)K^2)$ with $K=\Theta(x_{\max}y_{\max}+\frac{\ln(1/\epsilon)}{\ln(e+\frac{\ln(1/\epsilon)}{x_{\max}y_{\max}})})$.
\end{itemize}
Where $\alpha=e^{2x_{\max}y_{\max}}(2e^{x_{\max}y_{\max}/2}-1)^2$, $\hat{I}_{k,k'}=\sum_{i=1}^{N}I_k(\tilde{x}_i)I_{k'}(\tilde{x}_i')/(I_{k,\max}I_{k',\max}')\ket{i}\bra{i}$, $\hat{T}_{k,k'}=\sum_{j=1}^{N}T_k(\tilde{y}_j)T_{k'}(\tilde{y}_j')\ket{j}\bra{j}$,  and $\tilde{O}$ corresponds to a $O$ scaling up to a factor polylogarithmic in $K$.    
\end{theorem}

The proof is similar to the proof presented in the real case.

The QLT with complex coefficients can be  used to approximate the continuous Laplace transform as shown in Appendix \ref{sec: QLP and continuous}.

\section{Discussion}
\label{sec: discussion}

\paragraph{Non-unitary diagonal operators.}
The efficiency of the QLT relies on the quantum circuits associated with the non-unitary diagonal operators that encode the input data $\{x_i\}_{1\leq i\leq N}$ and $\{y_j\}_{1\leq j\leq N}$. Their implementation is not straightforward and depends directly on the structure of their diagonal coefficients. For an arbitrary $n$-qubit diagonal operator, exact implementations require an amount of resources that scales exponentially with $n$ in terms of either depth or ancilla qubits \cite{bullock2008asymptotically, sun2023asymptotically,zylberman2024efficient_bis}. Four cases with efficient implementations are referenced. The first one concerns sparse diagonal operators for which only $s$ of eigenvalues are non-vanishing. In this case, the block-encoding can be constructed with two $s$-sparse diagonal unitaries each controlled by the same ancilla qubit (see section 5 in \cite{zylberman2024efficient_bis}). The diagonal unitaries are themselves composed of $s$-multicontrolled phase gates which implement the eigenvalues one by one (see sequential decomposition in \cite{zylberman2024efficient_bis}). The second case of interest deals with diagonal operators that depend on functions $f$ with some smoothness properties. In this case, the eigenvalues of the diagonal operator are directly given by the values of the function at different positions $\hat{f}=\sum_x f(x)\ket{x}\bra{x}$. When the function $f$ can be expanded in a series, such as Fourier or Walsh-Hadamard: one can implement the associated series at a cost proportional to the number of terms implemented. If $f$ is continuously differentiable, a Walsh-Hadamard expansion of $f$ provides an $\epsilon$-approximation of the associated diagonal operator with complexity $O(||f'||_\infty/\epsilon)$ \cite{welch2014efficient,zylberman2024efficient}. If the eigenvalues depend on a periodic and analytic function that is Fourier expandable with an exponential convergence rate, the number of terms required to reach an accuracy $\epsilon>0$ is logarithmic in $1/\epsilon$, as is the circuit depth. To implement the Fourier series, one can use the generalized quantum signal processing protocol (GQSP) \cite{motlagh2024generalized}. The GQSP protocol implements polynomials of unitary operators. In this case, the Fourier series of the $n$-qubit diagonal operator corresponds directly to a polynomial of a tensor product of $n$ phase gates. This polynomial can be $(1,n,\epsilon)$-block-encoded with a circuit depth $O(\log(1/\epsilon))$ implying that the QLT has a polylogarithmic depth (this case is detailed in Appendix \ref{sec: QLP with Fourier}). The last method consists of expressing the eigenvalues as a closed-form operation. By using arithmetic routines, one can implement the associated diagonal operation with primitive quantum gates \cite{kashefi2002comparison,childs2017lecture}. The synthesis of efficient quantum circuits for unitary and non-unitary diagonal operators is an active area of research \cite{sun2023asymptotically,motlagh2024generalized,zylberman2024efficient_bis} and it is likely that more efficient methods will be developed in the coming years.

\paragraph{Applications.}
The QLT introduced in this work is not a quantum algorithm on its own but it can be used as a routine for different quantum algorithms targeting applications that leverage Laplace transforms. For instance, the $Z$-transform is a particular case of the discrete Laplace transform which is used in digital signal processing \cite{freeman1958bibliography,naredo2007z} and solving difference equations \cite{pospivsil2017representation,mahmudov2020delayed}. The $Z$-transform of a sequence $\{s(i)\}_{0\leq i \leq N-1}$ is defined by $Z\{s(i)\}(z)=\sum_{i=0}^{N-1} s(i)z^{-i}=\sum_{i=0}^{N-1} s(i)e^{-i\ln(z)}$. Computing $Z\{s(i)\}$ from a sequence encoded in a qubit state $\ket{s} \propto \sum_{i=0}^{N-1}s(i)\ket{i}$ at $N$ different $\{z_j\}_{0\leq j \leq N-1}\in \mathbb{R}^N$ is obtained using the QLT associated with the coefficients $x_i=i$ and $y_j=\ln(z_j)$:  $\ket{Z\{s(i)\}} \propto \widehat{QLT}\ket{s}$ up to normalization. Additionally, the $Z$-transform enables the creation of generating function associated with finite probability distributions encoded in a qubit state. Other examples directly involving a discrete Laplace transform arise in physics with the decay of a radioactive sample containing $N$ isotope species, in quantum Monte Carlo methods to compute a spectral function $A(\omega)$ from the imaginary-time Green function $G(\tau)$ or in statistical mechanics to compute the partition function $Z(\beta)$ at different inverse temperatures $\beta_j$ for systems with varying energy levels and multiplicities \cite{loh2023fast}.  

The continuous Laplace transform is studied more extensively than the discrete Laplace transform, and the QLT can approximate the continuous Laplace transform. When it is well-defined, the continuous Laplace of a function $f$ is  $\mathcal{L}\{f\}(z)=\int_0^{+\infty}f(t)e^{-zt}dt$. A way to approximate $\mathcal{L}\{f\}$ is to truncate and discretize the integral as $\mathcal{L}\{f\}(z)\simeq \sum_{i=0}^{N-1}e^{zt_i}f(t_i)t_{\max}/N$. In particular, if the values $f(t_i)$ are encoded in a qubit state $\ket{f}\propto \sum_{i=0}^{N-1}f(t_i)\ket{i}$, one can use the QLT to approximate the qubit state that encodes the continuous Laplace transform at various $z_j\in \mathbb{C}$: $\ket{\mathcal{L}\{f\}}\propto \sum_{j=0}^{N-1}\mathcal{L}\{f\}(z_j)\ket{j}\simeq \widehat{QLT}\ket{f}$. Similarly, the inverse Laplace transform can be approximated using a QLT, especially when the integration is given by a closed curve in $\mathbb{C}$. More details are given in Appendix \ref{sec: QLP and continuous}. 

\paragraph{Probability of success.}  One drawback of using a non-unitary operation $\hat{A}$ on quantum computers is that the block $\hat{U}_{\hat{A}}$ applied on a qubit $\ket{f}\ket{0}^{\otimes b}$ produces a state $\hat{A}\ket{f}\ket{0}^{\otimes b}+\ket{\phi_{\perp}}$ with $\ket{\phi_{\perp}}$ being orthogonal to the state $\hat{A}\ket{f}\ket{0}^{\otimes b}$. The amplitude of the target state is therefore smaller than $1$. Performing many non-unitary operations can lead to a prohibitively small probability of success. To circumvent this issue, one can performed amplitude amplification techniques  \cite{brassard2002quantum,berry2014exponential}. Another interesting case is when the probability of success converges toward a constant independent of $N$ and $\epsilon$. For instance, when the coefficients of the QLT and the qubit state depends on continuously differentiable functions $f,g,h$ defined on $[0,1]$ as $x_i=f(i/N)$, $y_j=g(j/N)$ for all $i,j=0,\hdots,N-1$ and $\ket{h}=(1/\|h\|_{2,N})\sum_{i=0}^{N-1}h(i/N)\ket{i}$, then the probability of success converges toward $e^{-\|f\|_{\infty}\|g\|_{\infty}}\int_0^1|\int_0^1 e^{f(x)g(y)}h(y)dy|^2dx/\int_0^1|h(y)|^2dy=\Theta(1)$.
 
\paragraph{Potential quantum advantage.} The performance of the QLT is presented in terms of size, depth and number of ancilla qubits but other metrics may be more relevant depending on hardware specifications. The depth is related to the circuit runtime but, in some cases, its scaling could be significantly affected by hardware constaints. For instance, a linear connectivity of the qubits prevents any double-logarithmic scaling of the quantum circuit since one qubit can "communicate" with another only after traversing all the qubits between them. Another constraint is the basis of universal primitive quantum gates permitted by a given quantum processor. In this work, we synthesized the quantum circuit in terms of single- and two-qubit gates without further specifications on the gate set. In practice, some specific quantum gates might be more difficult to implement than others.  For example, $T$ gates can be a limiting factor to achieve parallelization of quantum gates since the production of parallel $T$ gates is limited by the number of parallel magic-state factories that produce them.  These issues are common across quantum algorithms and routines and the full potential of quantum computers will only be realized with fault-tolerant digital quantum processors offering high connectivity and parallelized quantum gate implementation. In this case, the presented quantum circuits of the QLT can  offer a super-exponential or a double-exponential speedup in time compared to performing this same operation on a classical computer. However, this routine speedup is not an algorithmic speedup, as we are not solving any decision problem in this work.

\paragraph{} Additionally, we note that this research was developed independently of the very recent work \cite{an2024laplace}. In that paper, the authors introduce a quantum eigenvalue transformation based on a specific type of matrix Laplace transformation, enabling the implementation of certain classes of non-unitary operators up to some approximations. This approach is complementary to the one presented here, where we consider the Laplace transform as an operation on qubit states. 

\section{Conclusion}
This work introduces the quantum Laplace transform (QLT) operation, defined as $(\widehat{QLT})_{i,j}=e^{x_iy_j}/N$ with $x_i,y_j\in \mathbb{C}$. This operation corresponds to an $N\times N$ discrete Laplace transform. We demonstatre that the QLT can be approximated with an error of at most $\epsilon>0$ with $O(\log(1/\epsilon))$ diagonal operators and a quantum circuit of depth $\tilde{O}(\log(\log(N))\log(1/\epsilon))$, size $\tilde{O}(\log(N)\log(1/\epsilon))$, and using $O(\log(N)+\log(1/\epsilon))$ ancilla qubits. In many cases, the diagonal operators are implementable with shallow-depth quantum circuits. This leads to the  result that discrete Laplace transforms can be executed with exponentially fewer primitive operations on quantum computers than on classical computers, potentially achieving a double-exponential reduction in computational time. These constructions are the first in the literature to enable the implementation of the discrete Laplace transform on quantum computers. Notably, the QLT can also approximate both the continuous Laplace transform and its inverse on quantum computers.

Similar to the fast quantum Fourier transform \cite{cleve2000fast}, the QLT offers a routine speedup that could be leveraged in future research. Potential directions include developing quantum algorithms for problems whose classical solutions rely on Laplace transforms, exploring opportunities for algorithmic quantum advantage, and creating entirely new quantum algorithms based on the QLT. Futhermore, various classical methods for the inverse Laplace transform could be adapted to the quantum computing framework using the ideas presented here \cite{kuhlman2013review}.

\section*{Acknowledgements}
The author would like to thank F. Debbasch, B. Claudon, C. Feniou, L. Londeix Pagnard and M. Arsenault for their useful feedback on the form and content of this research.

\bibliography{main.bib}
\bibliographystyle{unsrt}

\appendix
\section{Additional lemmas and proofs}

\subsection{Proof of Lemma \ref{lemma: approx of exp} }
\label{Proof of Lemma approx exp}
The first inequality is a direct consequence of Taylor's theorem: For any $(K+1)$-continuously differentiable function $f$ defined on an interval $[a,b]$, one has:
\begin{equation}
    f(b)=\sum_{k=0}^{K}\frac{f^{(k)}(a)}{k!}(b-a)^k+\int_a^b\frac{f^{(K+1)}(s)}{K!}(b-s)^{K}ds.
\label{Taylor formula}
\end{equation}

The second inequality is inspired from equation (85) in \cite{berry2015hamiltonian} and equation (56) in \cite{gilyen2019quantum}. The proof uses the fact that $|T_k(y)|\leq1$ for all $k\in\mathbb{N}, y \in [-1,1]$, the equality $I_k(z)=(-\mathbf{i})^kJ_k(\mathbf{i}z), \forall z\in\mathbb{C},k\in \mathbb{N}$, where $J_k$ is the Bessel function of order $k$, $\mathbf{i}^2=-1$, and the bound $|J_k(z)|\leq |z/2|^ke^{|\Im(z)|}/k!$ valid for all $k\in\mathbb{N}^*$ and $z\in\mathbb{C}$, where $\Im(z)$ is the imaginary part of $z$ (Equation  9.1.62 in \cite{abramowitz1968handbook}).
\begin{equation}
\begin{split}
    |2\sum_{k=K+1}^\infty I_k(\tilde{x}_i)T_k(\tilde{y}_j)|&\leq2e^{x_{\max}y_{\max}}\sum_{k=K+1}^\infty \frac{1}{k!}|\frac{\tilde{x}_i}{2}|^k\\&\leq \frac{2e^{x_{\max}y_{\max}}}{(K+1)!}|\frac{\tilde{x}_i}{2}|^{K+1}\sum_{k=0}^\infty \frac{(K+1)!}{(K+1+k)!}|\frac{\tilde{x}_i}{2}|^k \\ &\leq \frac{2e^{x_{\max}y_{\max}}}{(K+1)!}|\frac{\tilde{x}_i}{2}|^{K+1}\sum_{k=0}^\infty(\frac{1}{2})^k 
\end{split}
\end{equation}
Last inequality is valid under the condition $K+1>|\tilde{x}_i|$ or $K+1>x_{\max}y_{\max}$.

\subsection{Proof of inequality \ref{inequality 2Dcase}}
\label{proof: 2Dcase}
The QLT with complex coefficients can be approximated with a double Chebyshev series as:
\begin{equation}
\begin{split}
    &|e^{x_iy_j}-\sum_{k=0}^K\sum_{k'=0}^{K}(2-\delta_k)(2-\delta_{k'})I_{k}(\tilde{x}_i)I_{k'}(\tilde{x}_i')T_k(\tilde{y}_j)T_{k'}(\tilde{y}_j')|\\& 
      \leq |e^{\tilde{x}_i\tilde{y}_j}(e^{\tilde{x}_i'\tilde{y}_j'}-\sum_{k'=0}^{K}I_{k'}(\tilde{x}_i')T_{k'}(\tilde{y}_j'))|+|(\sum_{k'=0}^{K}I_{k'}(\tilde{x}_i')T_{k'}(\tilde{y}_j'))(e^{\tilde{x}_i\tilde{y}_j}-\sum_{k=0}^{K}I_{k}(\tilde{x}_i)T_{k}(\tilde{y}_j))|
    \\
    &\leq e^{x_{\max}y_{\max}}(\frac{4e^{x_{\max}y_{\max}}}{(K+1)!}|\frac{\tilde{x}_i'}{2}|^{K+1})+e^{x_{\max}y_{\max}}(2e^{x_{\max}y_{\max}/2}-1)(\frac{4e^{x_{\max}y_{\max}}}{(K+1)!}|\frac{\tilde{x}_i}{2}|^{K+1})
    \\
    &\leq e^{\frac{5}{2}x_{\max}y_{\max}}\frac{8}{(K+1)!}|\frac{x_{\max}y_{\max}}{2}|^{K+1}
\end{split}
\end{equation}
which is valid for $K+1\ge x_{\max}y_{\max}$.

\subsection{Lemmas for parallelizing controls }
\label{sec:parallel control}

We recall the lemma II.I of \cite{zylberman2024efficient_bis}.
Consider a qubit in an arbitrary state $\ket{\psi}=\alpha\ket{0}+\beta\ket{1}$ and a register of $n_a-1$ zeroed ancilla qubit in state $\ket{0}^{\otimes (n_a-1)}$. The copy unitary $\hat{C}$ copies acts as:
\begin{equation}
\label{eq: copy}
\hat{C}(\alpha\ket{0}+\beta\ket{1}) \otimes \ket{0}^{\otimes (n_a-1)}=\alpha\ket{0}^{\otimes n_a}+\beta\ket{1}^{\otimes n_a}.
\end{equation}

\begin{lemma}
Let $n_a \in \mathbb{N}^*$, the “copy” unitary $\hat{C}$ acting on $n_a$ qubits can be implemented using $n_a-1$ CNOT gates with a depth $\lceil \log_2(n_a) \rceil$.
\label{copy lemma}
\end{lemma}
\begin{proof}
Note first that a set of $q$ CNOT gates applied to different qubits can be implemented in parallel with a circuit depth $1$. Define $k=\lceil \log_2(n_a) \rceil$. A direct recursion on $k'=1,...,k-1$ proves that parallel CNOT gates can double the number of “copied” qubit at each step, so each newly copied qubit can be used to copy another one at the next step as 
\begin{equation}
\begin{split}
(\alpha\ket{0}+\beta&\ket{1})\otimes \ket{0}^{\otimes (n_a-1)} \xrightarrow{\widehat{CNOT}_{0\rightarrow1}} (\alpha\ket{0}^{\otimes 2}+\beta\ket{1}^{\otimes 2})\otimes \ket{0}^{\otimes (n_a-2)} \\ & \xrightarrow{\widehat{CNOT}_{0\rightarrow2}\otimes \widehat{CNOT}_{1\rightarrow3}} (\alpha\ket{0}^{\otimes 4}+\beta\ket{1}^{\otimes 4})\otimes \ket{0}^{\otimes (n_a-4)} \\ & \vdots 
\\ & \xrightarrow{\otimes_{j=0}^{2^{k'}-1}\widehat{CNOT}_{j\rightarrow (j+2^{k'})}} \alpha\ket{0}^{\otimes2^{k'+1}}+\beta\ket{1}^{\otimes 2^{k'+1}}\otimes \ket{0}^{(n_a-2^{k'+1})},
\end{split}
\end{equation}

where $\widehat{CNOT}_{j\rightarrow (j+2^{k'})}$ is the CNOT gate controlled by the $j$-th qubit and applied on $(j+2^{k'})$-th zeroed qubit. After $k-1$ steps, the qubit state is $(\alpha\ket{0}^{\otimes2^{k-1}}+\beta\ket{1}^{\otimes 2^{k-1}})\otimes\ket{0}^{\otimes(n_a-2^{k-1})}$. A last layer of CNOT gates suffices to produce the state $\alpha\ket{0}^{\otimes n_a}+\beta\ket{1}^{\otimes n_a}$. 

\end{proof}

\begin{lemma} (Parallel control of one layer of quantum gates)
\label{lemma: parallel control}
Let $n\in \mathbb{N}^*, n_a\in \mathbb{N}^*$, $1\leq n_a \leq n$, $\{q_i\}_{1\leq i\leq n}$ and $\{a_i\}_{1\leq i\leq n_a}$ two sets of $n$ and $n_a$ qubits, $\hat{G}_i\in U(4)$ for $i=1,\hdots,n_a$ such that each $\hat{G}_i$ can be implemented on a distinct subset of the $\{q_i\}_{1\leq i\leq n}$  qubits in parallel of the others, $\hat{I}_q$ denotes the identity operator acting on the $\{q_i\}_{1\leq i\leq n}$ qubits and $C_{a_i}(\hat{U})=\hat{I}\otimes\ket{a_i=0}\bra{a_i=0}+\hat{U}\otimes\ket{a_i=1}\bra{a_i=1}$ the operator $\hat{U}$ controlled by the $a_i$ qubit, the operator $C_{a_1}(\bigotimes_{i=1}^{n_a}\hat{G}_i)=\hat{I}_q\otimes\ket{a_1=0}\bra{a_1=0}+\bigotimes_{i=1}^{n_a}\hat{G}_i\otimes\ket{a_1=1}\bra{a_1=1}$ verifies
\begin{equation}
    C_{a_1}(\bigotimes_{i=1}^{n_a}\hat{G}_i)=\bra{0}^{\otimes(n_a-1)} (\hat{C}(\bigotimes_{i=1}^{n_a}C_{a_i}(\hat{G}_i))\hat{C})\ket{0}^{\otimes(n_a-1)} 
\end{equation}
with $\hat{C}$ the copy unitary defined Eq \ref{eq: copy} and therefore can be $(1,n_a-1,0)$-block-encoded with a circuit of depth $O(\log(n_a))$ and size $O(n_a)$.

\end{lemma}
\begin{proof}
    We consider the $2n$-qubit operation $\hat{C}(\bigotimes_{i=1}^{n_a}C_{a_i}(\hat{G}_i))\hat{C}$.
\begin{equation}
\begin{split}
     &\hat{C}(\bigotimes_{i=1}^{n_a}C_{a_i}(\hat{G}_i))\hat{C}\ket{q}\otimes(\alpha\ket{0}+\beta \ket{1})\otimes\ket{0}^{\otimes(n_a-1)} 
     =\hat{C}(\bigotimes_{i=1}^{n_a}C_{a_i}(\hat{G}_i))\ket{q}\otimes(\alpha\ket{0}^{\otimes n_a}+\beta \ket{1}^{\otimes n_a})\\
     &=\hat{C}(\alpha\ket{q}\otimes\ket{0}^{\otimes n_a}+\beta  (\bigotimes_{i=1}^{n_a}\hat{G}_i)\ket{q}\otimes \ket{1}^{\otimes n_a} ) 
     = (\alpha\ket{q}\otimes\ket{0}+\beta  (\bigotimes_{i=1}^{n_a}\hat{G}_i)\ket{q}\otimes \ket{1})\otimes \ket{0}^{\otimes (n_a-1)} \\
     & = C_{a_1}(\bigotimes_{i=1}^{n_a}\hat{G}_i) \ket{q}\otimes(\alpha\ket{0}+\beta \ket{1})\otimes\ket{0}^{\otimes(n_a-1)} 
\end{split}
\end{equation}
Each $\hat{C}$ is equal to $n_a$ CNOT gates arranged in $\lceil\log(n_a)\rceil$ layers (see lemma \ref{copy lemma}) and the $(\bigotimes_{i=1}^{n_a}C_{a_i}(\hat{G}_i))$ operations is composed of one layer of $n_a$ two-qubit and three-qubit gates. The three qubit gates are decomposable into $O(1)$ one and two-qubit quantum gates using a set of universal quantum gates. Therefore,  the operation $ C_{a_1}(\bigotimes_{i=1}^{n_a}\hat{G}_i)$ can be $(1,n_a,0)$-block-encoded with a circuit of depth $O(\log(n_a))$ and size $O(n_a)$
\end{proof}

\begin{lemma} (Control-unitary with same depth scaling)
\label{lemma : parallel control of one unitary}
Let $n \in \mathbb{N}^*$, $n_a \in \mathbb{N}^*$ and $\hat{U}$ be a $n$-qubit unitary implementable with a quantum circuit of size $O(s(n))$ and depth $O(d(n))$. Then, the one-control-$\hat{U}$ operation $C_{a_1}(U)=\hat{I}_n\otimes\ket{a_1=0}\bra{a_1=0}+\hat{U}\otimes\ket{a_1=1}\bra{a_1=1}$ can be $(1,n_a-1,0)$-block-encoded with a quantum circuit of size $O(s(n)+n_a)$ and depth $O(\lceil n/n_a\rceil d(n)+ \log(n_a))$.
\end{lemma}

\begin{proof}
    The $n$-qubit unitary $\hat{U}$ with depth $d$ can be represented as a product of $d$ layers of primitive quantum gates $\hat{U}=\prod_{i=1}^d (\bigotimes_{j=1}^{n_i} \hat{U}_{i,j})$ where $n_i$ is the number of primitive quantum gates in the $i-$th layer $n_i\leq n$ and the $\hat{U}_{i,j}\in U(4)$ are single- or two-qubit unitaries.
    Each layer can be decomposed into $\lceil n_i/n_a\rceil$ layers of at most $n_a$  gates: $\bigotimes_{j=1}^{n_i}\hat{U}_{i,j}=\prod_{k=0}^{\lceil n_i /n_a \rceil-1}(\bigotimes_{j=1}^{n_{i,k}}\hat{U}_{i,j,k})$ with $\sum_{k=0}^{\lceil n_i /n_a \rceil-1}n_{i,k}=n_i$, $n_{i,k}\leq n_a$ for all $k=0,\hdots,\lceil n_i /n_a \rceil-1$ and $\hat{U}_{i,j,k}=\hat{U}_{i,j+\sum_{k'=0}^kn_{i,k'}}$. Then, using the Lemma \ref{lemma: parallel control} on parallel controls for one layer of quantum gates, one has:

\begin{equation}
\begin{split}
    C_{a_1}(\hat{U})&=C_{a_1}(\prod_{i=1}^d \prod_{k=0}^{\lceil n_i /n_a \rceil-1}(\bigotimes_{j=1}^{n_{i,k}}\hat{U}_{i,j,k}))
    =\prod_{i=1}^d \prod_{k=0}^{\lceil n_i /n_a \rceil-1}C_{a_1}(\bigotimes_{j=1}^{n_{i,k}}\hat{U}_{i,j,k}))
    \\&=\prod_{i=1}^d \prod_{k=0}^{\lceil n_i /n_a \rceil-1}\hat{C} \bigotimes_{j=1}^{n_{i,k}}C_{a_j}(\hat{U}_{i,j,k}) \hat{C} =\hat{C}(\prod_{i=1}^d \prod_{k=0}^{\lceil n_i /n_a \rceil-1} \bigotimes_{j=1}^{n_{i,k}}C_{a_j}(\hat{U}_{i,j,k}) ) \hat{C}
\end{split}
\end{equation}
where $\hat{C}$ is the copy unitary defined Eq\ref{eq: copy} acting on $n_{a}$ qubits. The $\hat{C}_{a_j}(\hat{U}_{i,j,k})$ are two-qubit or three-qubits gates. For the three qubit gates, canonical methods exist to decompose them into $O(1)$ one-qubit and two-qubit gates.   Therefore, the operation $C_{a_1}(\hat{U})$ can be $(1,n_a,0)$-block-encoded with a quantum circuit of size $O(n_a+s(n))$ and depth $O(\lceil n/n_a\rceil d(n)+\log(n_a))$.
\end{proof}

In particular, when $n_a=n$, the control-$\hat{U}$ operation can be $(1,n-1,0)$-block-encoded with a circuit of depth $O(d(n)+\log(n))$.

\section{Logarithmic-depth circuit for non-unitary diagonal operator}
\label{sec: QLP with Fourier}

In the past few years, different methods of quantum signal processing have been developed to implement polynomials of unitary operators \cite{Low2017,gilyen2019quantum,motlagh2024generalized}. It has been highlighted that Fourier series associated to diagonal operators can be block-encoded using these protocols \cite{motlagh2024generalized}.

Consider a real-valued function $g$ defined on $[0,1]$ and the following non-unitary diagonal operator:
\begin{equation}
\begin{split}
    \hat{D}=\sum_{j=0}^{N-1}g(j/N)\ket{j}\bra{j}
\end{split}
\end{equation}

When the function $g$ can be represented exactly by a Fourier series, one can approximate $g$ using a truncated Fourier series:
\begin{equation}
\begin{split}
\tilde{g}_M(x)=\sum_{k=-M}^M  a_{k}^g e^{2 i\pi kx}
\end{split}
\end{equation}
with $a_{k}^g=\int_{0}^1g(x)e^{-2i\pi kx}dx$ the $k$-th Fourier coefficient of $g$. In particular, if $g$ is analytic and periodic, and there exist two constants $C>0$ and $R>1$ such that $\forall M \in \mathbb{N}$, $\|\partial_{x}^{(M)}g\|_{\infty}\leq C M!/R^M$, the Fourier series $\tilde{g}_M$ converges exponentially to $g$ as $M$ increases.

The associated diagonal operator $\hat{D}_M=\sum_{j=0}^{N-1}\tilde{g}_M(j/N)\ket{j}\bra{j}$ can be recast as a polynomial of the simple operator $\hat{U}_{\omega}=\sum_{j=0}^{N-1}e^{2i\pi j/N}\ket{j}\bra{j}$ \cite{motlagh2024generalized}:
\begin{equation}
    \hat{D}_M=\sum_{k=-M}^M  a_{k}^g (e^{2 i\pi \hat{x}})^k=e^{-2 i\pi M\hat{x}}\sum_{k=0}^{2M}  a_{k-M}^g (e^{2 i\pi \hat{x}})^k=e^{-2 i\pi M\hat{x}}P(\hat{U}_\omega)
\end{equation}
with $\hat{x}=\sum_{j=0}^{N-1}(j/N)\ket{j}\bra{j}$, $P(X)=\sum_{k=0}^{2M}  a_{k-M}^gX^k$. 

The generalized quantum signal processing (GQSP) protocol enables the block-encoding of polynomial of unitaries with a cost proportional to the degree $d$ of the polynomial. The associated quantum circuit is composed of $d$ single qubit rotations and $d$ one-controlled $\hat{U}_{\omega}$. The angles of the one-qubit rotations that produce the polynomial $P$ require to be computed with a specific classical optimization \cite{motlagh2024generalized}, or Fourier-based classical integration \cite{berntson2024complementary}. The GQSP method associated with the parallelization of controls lemmas provided in Appendix \ref{sec:parallel control}, and the exponential convergence of the Fourier series, provides an efficient logarithmic-depth method to block-encode non-unitary diagonal operators. This result is summarized in the following lemma.

\begin{lemma} Logarithmic-depth quantum circuits for Fourier series approximation of diagonal operators

Let $n\in\mathbb{N}^*$, $N=2^n$, $C>0$, $R>1$, $\hat{D}=\sum_{j=0}^{N-1}g(j/N)\ket{j}\bra{j}$ with $g:[0,1] \rightarrow\mathbb{R}$ a periodic and analytic function satisfying $\forall M \in \mathbb{N}$, $\|\partial_{x}^{(M)}g\|_{\infty}\leq C M!/R^M$ with $\|g\|_{\infty}=\max_{x\in[0,1]}|g(x)|$. Then, the diagonal operator $\hat{D}$  can be $(\|g\|_{\infty},n,\epsilon)$-block-encoded using a quantum circuit of depth $O(\log(1/\epsilon))$ and size $O(n\log(1/\epsilon))$.
\end{lemma}
\begin{proof}

First, one can show, using similar computations as those presented in the Appendix of \cite{childs2022quantum}, that for all $M\in\mathbb{N}$,:
\begin{equation}
     \|g-\tilde{g}_M\|_\infty \leq2\frac{\|\partial_{x}^{(M)}g\|_{\infty}}{(2\pi)^{M}(M-1) M^{(M-1)}}
\label{Fourier convergence}
    \end{equation}

Then, there exist two constants $C>0$ and $R>1$ such that for all $M\in\mathbb{N}$, $\|\partial_{x}^{(M)}g\|_{\infty}\leq C M!/R^M$:

\begin{equation}
     \|g-\tilde{g}_M\|_\infty \leq2C\frac{M!}{(2\pi R)^{M}(M-1) M^{(M-1)}} \leq 2C\frac{\sqrt{2\pi M^3} }{(2\pi e R)^{M}(M-1)} e^{\frac{1}{12M}}=\epsilon_M
    \end{equation}
using Stirling's inequality $M!\leq \sqrt{2\pi M}(M/e)^Me^{\frac{1}{12M}}$. It suffices to choose $M$ scaling as $\Theta(\log(1/\epsilon))$ to get an asymptotic $\epsilon$-approximation of $g$ using its truncated Fourier series $\tilde{g}_M$.

Then, the generalized quantum signal processing protocol \cite{motlagh2024generalized}  allows the implementation of polynomial of operators without additional restrictions on the polynomial other than $|P(e^{i\theta})|\leq1$ for $\theta\in[0,2\pi[$. We refer to Corollary 14 of \cite{motlagh2024generalized}: Let $f (x) = \sum_{k=0}^{d}\alpha_ke^{2i\pi kx/N}$ be a function such that $\|f\|_{\infty}\leq 1$, it is possible to $(1,1,0)$-block encode the $n$-qubit operator $\sum_{k=0}^d f(x)\ket{x}\bra{x}$ with $O(dn)$ single- and two-qubit gates.
In our case, consider the normalized function $\tilde{\tilde{g}}_M=\tilde{g}_M/(\|g\|_{\infty}+\epsilon_M)$ such that $\|\tilde{\tilde{g}}_M\|_{\infty}\leq 1$, and:
\begin{equation}
    \|\frac{g}{\|g\|_{\infty}}-\tilde{\tilde{g}}_M\|_{\infty}\leq\frac{2\epsilon_M}{\|g\|_{\infty}}
\end{equation}
The GQSP protocol, using the polynomial $P_M(X)=\sum_{k=0}^{2M}(\alpha_{k-M}/(\|g\|_{\infty}+\epsilon_M))X^k$ and the unitary $\hat{U}_{\omega}=\sum_{j=0}^{N-1}e^{2i\pi j/N}\ket{j}\bra{j}$, produces a $(1,1,0)$-block encoding $\hat{W}$ of $\tilde{D}_M=\sum_{k=0}^d \tilde{\tilde{g}}_M(x)\ket{x}\bra{x}$. This operator $\hat{W}$ is a $(\|g\|_{\infty},1,2\epsilon_M)$-block-encoding of $\hat{D}$: 
\begin{equation}
\|\hat{D}-\|g\|_{\infty}(\bra{0}\otimes \hat{I}_n)\hat{W}(\ket{0} \otimes \hat{I}_n)\|_2\leq\|g\|_{\infty}\|\frac{g}{\|g\|_{\infty}}-\tilde{\tilde{g}}_M\|_{\infty}\leq2\epsilon_M
\end{equation}

The associated quantum circuit has a size and depth of $O(Mn)$, using one ancilla qubit. The factor $n$ in the depth arises from the one-controlled $\hat{U}_{\omega}$. The operator $\hat{U}_{\omega}$ corresponds to $n$ one-qubit phase gates. To control the $n$ one-qubit gates with a depth of $O(\log(n))$, one can use $n-1$ ancilla qubits and Lemma \ref{lemma: parallel control} to parallelize the controls. The diagonal operation can, therefore, be $(\|g\|_{\infty},n,\epsilon)$-block-encoded with a circuit depth of $O(\log(n)\log(1/\epsilon))$ and a size of $O(n\log(1/\epsilon))$.

\end{proof}

\section{Relation between the quantum Laplace transform, the continuous Laplace Transform and its inverse}
\label{sec: QLP and continuous}
The continuous Laplace and its inverse are more commonly used in the literature than the discrete Laplace transform. Their implementation on classical and quantum computers requires some approximations. In the following, we consider the problem of preparing a qubit state that encodes the continuous Laplace transform of a function $f$ from a qubit state encoding the values of $f$. The approach is straightforward and consists of truncating and discretizing the integral of the continuous Laplace transform. The associated operation is directly given by the quantum Laplace transform and we estimate the associated complexity.
Consider the continuous Laplace transform of a real-valued function $f:\mathbb{R_+}\rightarrow \mathbb{R}$, defined as:
\begin{equation}
    \mathcal{L}\{f\}(\sigma,\omega)=\int_0^{+\infty}e^{-(\sigma+\mathbf{i}\omega)t}f(t)dt
\end{equation}
with $\sigma,\omega\in\mathbb{R}$ and $\mathbf{i}^2=-1$.
The integral exists for $\sigma>a$ if \cite{honig1984method}.:
\begin{itemize}
\item (i) $f$ is locally integrable,
\item (ii)  there exist a $t_0\ge0$ and $k,a\in\mathbb{R}$ such that 
\begin{equation}
    |f(t)|\leq ke^{at}, \text{ } \forall t\ge t_0,
\end{equation}
\item (iii) for all $t\in \mathbb{R}_+$, there is a neighborhood in which $f$ is of bounded variation.
\end{itemize}

Consider the $n$-qubit vector whose components are directly given by $f$: 
\begin{equation}
    \ket{f}=\frac{1}{\|f\|_{2,N}}\sum_{j=0}^{N-1}f(t_j)\ket{j}
\end{equation}
with $\|f\|_{2,N}=\sqrt{\sum_{j=0}^{N-1}|f(t_j)|^2}$ ensuring normalization.

We show, starting from $\ket{f}$, how to implement an $\epsilon$-approximation in vector $2$-norm of the qubit state encoding the Laplace transform of $f$ at different $\{ \sigma_j\}_{0\leq j \leq N-1} \in \mathbb{R}^N$ and $\{ \omega_j\}_{0\leq j \leq N-1} \in \mathbb{R}^N$: 
\begin{equation}
    \ket{  \mathcal{L}\{f\} }=\frac{1}{\|  \mathcal{L}\{f\}\|_{2,N}}\sum_{j=0}^{N-1}  \mathcal{L}\{f\}(\sigma_j,\omega_j)\ket{j}
\end{equation}
with $\|  \mathcal{L}\{f\}\|_{2,N}=\sqrt{\sum_{j=0}^{N-1}|  \mathcal{L}\{f\}(\sigma_j,\omega_j)|^2}$

Two approximations are performed to approximate the values of the Laplace transform $  \mathcal{L}\{f\}(\sigma_j,\omega_j)$, the first one is the truncation of the integral and the second one is its discretization. When $f$ is of exponential order (condition (ii)), the integral can be $\epsilon$-approximated with a truncation parameter scaling with $\epsilon$ as $O(\log(1/\epsilon))$:

\begin{lemma}(Truncation error of the integral)
\label{lemma: Truncation error of the integral}

    Let $\epsilon\in\mathbb{R}_+$, $f$ be a real-valued function satisfying conditions (i),(ii),(iii) stated above, with parameters $t_0\in \mathbb{R}_+$, $k,a\in \mathbb{R}$ and $\omega\in\mathbb{R}$, $\sigma\in[\sigma_{\min},\sigma_{\max}]$ where $\sigma_{\min}>a$. Then, the Laplace transform $  \mathcal{L}\{f\}(\sigma,\omega)$ of $f$ can be $\epsilon$-approximated by:
\begin{equation}
    \int_0^Me^{-(\sigma+\mathbf{i}\omega)}f(t)dt
\end{equation}
with 
\begin{equation}
    M=\max(t_0,\frac{1}{\sigma_{\min}-a}\log(\frac{k}{\epsilon(\sigma_{\min}-a)}))
\label{Eq: def of M}
\end{equation}
\begin{proof}

\begin{equation}
\begin{split}
   & |\int_0^{+\infty}e^{-(\sigma+\mathbf{i}\omega)t}f(t)dt-\int_0^Me^{-(\sigma+\mathbf{i}\omega)t}f(t)dt|=|\int_M^{+\infty}e^{-(\sigma+\mathbf{i}\omega)t}f(t)dt| \\ &\leq k \int_M^{+\infty}e^{-(\sigma-a) t}dt
     \leq \epsilon
\end{split}
\end{equation}
\end{proof}

\end{lemma}

The integral $\int_0^M e^{-(\sigma+i\omega)t}f(t)dt$ can be numerically estimated by using brute force quadrature based on step functions. When the integrand is well-behaved (continuously differentiable with bounded variations), the integral can be $\epsilon$-approximatedl by a Riemann sum, with the number of interval scaling as $O(M^2/\epsilon)$: 
\begin{lemma} (Discretization error of the integral)
\label{lemma: Discretization error of the integral}

Let $M\in\mathbb{R}_+^*$, $N\in \mathbb{N}^*$, $\sigma,\omega \in \mathbb{R}$ and $f:[0,M]\rightarrow \mathbb{R}$ be a continuously-differentiable function. Then,

\begin{equation}
\| \int_0^{M}e^{-(\sigma+\mathbf{i} \omega)t}f(t)dt - \frac{M}{N}\sum_{i=1}^{N} e^{- (\sigma+\mathbf{i} \omega)t_i} f(t_i)\| \leq \frac{M^2}{2N} (\sqrt{\sigma^2+\omega^2}\max_{t\in[0,M]} |e^{-\sigma t} f(t)|+\max_{t\in[0,M]} |e^{-\sigma t} f'(t)|)
\end{equation}
with $t_i=(i-1)/N$ for all $i=0,\hdots,N$.

\end{lemma}

The following theorem summarizes the application of the previous lemmas for approximating the qubit state $\ket{  \mathcal{L}\{f\}}$.

\begin{theorem} (Continuous Laplace transform)
\label{thm: continuous laplace transform}

Let $\epsilon>0$, $n\in \mathbf{N}^*$, $N=2^n$,  $f:\mathbb{R_+}\rightarrow \mathbb{R}$ a continuously differentiable function satisfying conditions (i), (ii), (iii)  and $f'$ being also of exponential type, i.e., $f(t)\leq ke^{at}$ and $f'(t)\leq k' e^{a't}$ for $t \ge t_0$ and $k,k',a,a' \in \mathbb{R}_+$. We assume that the Laplace transform $  \mathcal{L}\{f\}$ of $f$ is continuously differentiable with bounded variations on a smooth curve $\mathcal{C}=\{z(s)\in \mathbb{C},  s\in[0,1] \}$ of bounded length.  We define $\sigma_j=\Re(z(j/N))>\max(a,a')$ and $\omega_j=\Im(z(j/N))$ for $j=0,\hdots,N-1$. Additionally, we also assume $N\ge M^2/\epsilon$ with $M$ defined in Equation \ref{Eq: def of M}.

Then, the $n$-qubit state $ \ket{  \mathcal{L}\{f\}}$, which encodes the Laplace transform of $f$ on the curve $\mathcal{C}$ at the $\{ \sigma_j \}_{1\leq j \leq N}$ and $\{ \omega_j\}_{1\leq j \leq N}$, can be prepared from $  \ket{f}=\frac{1}{\|f\|_{2,N}}\sum_{i=0}^{N-1}f(i\Delta t)\ket{i}$, where $\Delta t=M/N$, with an error bounded as $O(\epsilon)$ using the QLT associated to the $x_i=i\Delta t$ and $y_j=\sigma_j+\mathbf{i}\omega_j$ coefficients once. The probability of success is lower bounded as $\Omega(\epsilon^L)$, where $L=(a+2\max_{s\in[0,1]}|z(s)|)/(\min_{s\in[0,1]} \Re(z(s))-a)\ge 2$. 

\end{theorem}

\begin{proof}
First, note that, if the error in the implementation of the QLT is $\epsilon/2$, then the QLT can be $(e^{x_{\max}y_{\max}},b,\epsilon/2)$-block-encoded by a unitary $\hat{U}_{QLT}$ such that $\hat{U}_{QLT}\ket{f}\ket{0}^{\otimes b}=e^{-x_{\max}y_{\max}}\widehat{QLT}\ket{f}\ket{0}^{\otimes b}+\ket{\phi}_\perp$. Here, $x_{\max}=M(N-1)/N$, $y_{\max}=\max_{0\leq j \leq N-1} |\sigma_j+\mathbf{i}\omega_j|$ and $\ket{\phi}_\perp$ is a vector orthogonal to $\widehat{QLT}\ket{f}\ket{0}^{\otimes b}$. The probability of measuring the ancilla qubits in state $\ket{0}^{\otimes b}$ is given by:
\begin{equation}
\begin{split}
P=\|e^{-x_{\max}y_{\max}}\widehat{QLT}\ket{f}\|_2^2=\frac{e^{-2x_{\max}y_{\max}}}{M^2\|f\|_{2,N}^2}\sum_{j=0}^{N-1}|\frac{M}{N}\sum_{i=0}^{N-1}e^{-(\sigma_j+\mathbf{i}\omega_j)t_i}f(t_i)|^2
\end{split}
\end{equation}
with $t_i=iM/N$.

The smooth curve $\mathcal{C}$ on which the Laplace transform of $f$ is computed is determined by a bounded differentiable function $z:[0,1]\rightarrow \mathbb{C}$, implying that the coefficients  $\sigma_j=\Re(z(j/N))$ and $\omega_j=\Im(z(j/N))$ are bounded by a constant independent of $N$. Then, in the large $N$ limit, the sum over $j$ and $i$ converges to a constant integral as:
\begin{equation}
\frac{1}{N}\sum_{j=0}^{N-1}|\frac{M}{N}\sum_{i=0}^{N-1}e^{-(\sigma_j+\mathbf{i}\omega_j)t_i}f(t_i)|^2=\Theta(\int_0^1 |\int_0^M e^{-z(s)t}f(t)dt|^2 ds )
\end{equation}
and similarly, one can proove that, for a non-zero function $f$,:
\begin{equation}
\frac{N}{M\|f\|_{2,N}^2}=\Theta(\frac{1}{\int_0^M|f(t)|^2 dt})
\end{equation}
Therefore:
\begin{equation}
P=\Theta( e^{-2M\max_{s\in [0,1]} |z(s)|} \frac{\int_0^1 |\int_0^M e^{-z(s)t}f(t)dt|^2 ds}{M \int_0^M|f(t)|^2 dt})
\end{equation}
The term $\int_0^1 |\int_0^M e^{-z(s)t}f(t)dt|^2 ds$ converges to $\int_0^1 |\int_0^{+\infty} e^{-z(s)t}f(t)dt|^2 ds$ in the large $M$ limit while the term $\int_0^M|f(t)|^2 dt=\Omega(e^{-aM})$ since $f$ is of exponentially type. Thus, $P=\Omega(e^{-(a+2\max_{s\in [0,1]} |z(s)|)M}/M)$.

In the case of measuring the ancilla in the right state $\ket{0}^{\otimes b}$, the $n$-qubit state becomes $\widehat{QLT}\ket{f}/\|\widehat{QLT}\ket{f} \|_{2,N}$. In the following, we show that this state is close to the target state $\ket{  \mathcal{L}\{f\}}$. First, we define:
\begin{equation}
\begin{split}
\ket{f_1}&=\frac{\sum_{j=0}^{N-1}\int_0^Me^{-(\sigma_j+\mathbf{i}\omega_j)t}f(t)dt \ket{j}}{\|\sum_{j=0}^{N-1}\int_0^Me^{-(\sigma_j+\mathbf{i}\omega_j)t}f(t)dt \ket{j}\|_{2,N} }\\
\end{split}
\end{equation}
such that
\begin{equation}
\begin{split}
\| \frac{\widehat{QLT}\ket{f}}{\|\widehat{QLT}\ket{f} \|_{2,N} } - \ket{  \mathcal{L}\{f\}}\|_{2,N} \leq \| \frac{\widehat{QLT}\ket{f}}{\|\widehat{QLT}\ket{f} \|_{2,N} } - \ket{f_1}\|_{2,N}+\|\ket{f_1}-\ket{  \mathcal{L}\{f\}}\|_{2,N}
\end{split}
\end{equation}

The first term can be bounded as:
\begin{equation}
\begin{split}
 \| \frac{\widehat{QLT}\ket{f}}{\|\widehat{QLT}\ket{f} \|_{2,N} } - \ket{f_1}\|_{2,N} &\leq 2\frac{\|\sum_{j=0}^{N-1}(\frac{M}{N}\sum_{i=0}^{N-1}e^{-(\sigma_j+\mathbf{i}\omega_j)t_i}f(t_i)- \int_0^Me^{-(\sigma_j+\mathbf{i}\omega_j)t}f(t)dt) \ket{j}\|_{2,N}}{\|\sum_{j=0}^{N-1}\int_0^Me^{-(\sigma_j+\mathbf{i}\omega_j)t}f(t)dt \ket{j}\|_{2,N}} \\
\end{split}
\end{equation}
with $\sqrt{\sigma_j^2+\omega_j^2}\leq \max_{s\in[0,1]}|z(s)|$, and $f$, $f'$ being of exponential type, there exist $t_0\in\mathbf{R}$ such that $f(t)\leq ke^{at}$ and $f'(t)\leq k' e^{a't}$ for $t \ge t_0$, where $k,k',a,a' \in \mathbb{R}_+$, implying there exist two constants $K,K' \in \mathbb{R}_+$ independent of $N,M,j$ such that $\max_{t\in \mathbb{R}}|e^{-\sigma_jt}f(t)|\leq K$ and  $\max_{t\in \mathbb{R}}|e^{-\sigma_jt}f'(t)|\leq K'$. Thus, using lemma \ref{lemma: Discretization error of the integral}, the numerator is bounded by $K''M^2/\sqrt{N}$ with $K''\in  \mathbb{R}_+$ a constant independent of $N$ and $M$. Additionally, thanks to the fact that $  \mathcal{L}\{f\}$ is continuously differentiable with bounded variations on $\mathcal{C}$, we have $\|\sum_{j=0}^{N-1}\int_0^Me^{-(\sigma_j+\mathbf{i}\omega_j)t}f(t)dt \ket{j}\|_{2,N}=\Theta(\sqrt{N})$. Therefore,  $\| \frac{\widehat{QLT}\ket{f}}{\|\widehat{QLT}\ket{f} \|_{2,N} } - \ket{f_1}\|_{2,N}=O(M^2/N) $

Then, for the second term of the inequality, one has:
\begin{equation}
\begin{split}
\|\ket{f_1}-\ket{  \mathcal{L}\{f\} }\|_{2,N} & \leq  2\frac{\|\sum_{j=0}^{N-1}( \int_0^{+\infty}e^{-(\sigma_j+\mathbf{i}\omega_j)t}f(t)dt- \int_0^Me^{-(\sigma_j+\mathbf{i}\omega_j)t}f(t)dt) \ket{j}\|_{2,N}}{\|\sum_{j=0}^{N-1}\int_0^Me^{-(\sigma_j+\mathbf{i}\omega_j)t}f(t)dt \ket{j}\|_{2,N}}
\end{split}
\end{equation}
which, using Lemma \ref{lemma: Truncation error of the integral}, can be bounded independently of $N$ as:
\begin{equation}
\begin{split}
\|\ket{f_1}-\ket{F}\|_{2,N} = O( \max_{s\in[0,1]} e^{-(\Re (z(s))-a)M})
\end{split}
\end{equation}
Finally, for $M=O(\log(1/\epsilon))$, one has:
\begin{equation}
\begin{split}
\| \frac{\widehat{QLT}\ket{f}}{\|\widehat{QLT}\ket{f} \|_{2,N} } - \ket{  \mathcal{L}\{f\}}\|_{2,N} =O(\epsilon)
\end{split}
\end{equation}
and $P=\Omega(\epsilon^K)$ with $K=(a+2\max_{s\in[0,1]}|z(s)|)/(\min_{s\in[0,1]} \Re(z(s))-a)\ge 2$.
\end{proof}

This result enables the approximation of the continuous Laplace transform.  The main drawback arises from the truncation parameter $M=O(\log(1/\epsilon))$ which grows asymptotically towards $+\infty$, and from the normalization factor of the QLT $e^{\|z\|_{\infty}M}$, which scales as $O(1/\epsilon^{L'})$, where $L'>0$. Amplitude amplification techniques may be employed to improve the probability of success to $\Theta(1)$ \cite{brassard2002quantum,berry2014exponential}. Other discretization and truncation techniques may be explored to improve these scalings \cite{moreno2008implementation}.

In the case of the inverse Laplace transform, $f(t)=(e^{\sigma t}/(2\pi))\int_{-\infty}^{+\infty}\mathcal{L}\{f\}(\sigma,\omega)e^{\mathbf{i}\omega t}d\omega$, we note that similar truncation and discretization enable the approximation of the inverse Laplace transform using a QLT. If one has prepared a qubit state $\ket{ \mathcal{L}\{f\}}$  encoding the values of the Laplace transform $\mathcal{L}\{f\}$ on a specific curve $\mathcal{C}$ of the complex plane, it is possible to approximate the qubit state $\ket{f} \propto \sum_{j=0}^{2^n-1}\int_{\mathcal{C}}e^{z t_j}  \mathcal{L}\{f\}(z)dz \ket{j}$ using the QLT in a way similar to the method introduced for Theorem \ref{thm: continuous laplace transform}. 

\end{document}